\documentclass[fleqn,usenatbib]{mnras}

\usepackage{newtxtext,newtxmath}
\usepackage{xcolor}
\usepackage{makecell}
\usepackage{siunitx}
\usepackage{booktabs,caption}
\usepackage[flushleft]{threeparttable}

\usepackage[T1]{fontenc}

\DeclareRobustCommand{\VAN}[3]{#2}
\let\VANthebibliography\thebibliography
\def\thebibliography{\DeclareRobustCommand{\VAN}[3]{##3}\VANthebibliography}

\usepackage{graphicx}	% Including figure files
\usepackage{amsmath}	% Advanced maths commands
\usepackage{multirow}   % multi-row
\usepackage{makecell}

\title[Environmental Effects on AGN activity]{Environmental Effects on AGN activity via Extinction-free Mid-Infrared Census}

\author[D. J. D. Santos et al.]{Daryl Joe D. Santos$^{1}$\thanks{E-mail: daryl$\_$santos@gapp.nthu.edu.tw},
Tomotsugu Goto$^{1}$,
Seong Jin Kim$^{1}$,
Ting-Wen Wang$^{1}$, 
\newauthor
Simon C.-C. Ho$^{1}$, 
Tetsuya Hashimoto$^{1,2,3}$, 
Ting-Chi Huang$^{4,5}$,
Ting-Yi Lu$^{1}$,
\newauthor
Alvina Y. L. On$^{1,2,6}$,
Yi-Hang Valerie Wong$^{1}$,
Tiger Yu-Yang Hsiao$^{1}$,
Agnieszka Pollo$^{7,8}$,
\newauthor
Matthew A. Malkan$^{9}$,
Takamitsu Miyaji$^{10}$,
Yoshiki Toba$^{11,12,13}$,
Ece Kilerci-Eser$^{14}$,
\newauthor
Katarzyna Ma\l{}ek$^{7,15}$,
Ho Seong Hwang$^{16,17}$,
Woong-Seob Jeong$^{18,19}$,
Hyunjin Shim$^{20}$,
\newauthor
Chris Pearson$^{21,22,23}$,
Artem Poliszczuk$^{7}$, and
Bo Han Chen$^{24}$
\\
$^{1}$Institute of Astronomy, National Tsing Hua University, No. 101, Section 2, Kuang-Fu Road, Hsinchu City 30013, Taiwan \\
$^{2}$Centre for Informatics and Computation in Astronomy (CICA), National Tsing Hua University, 101, Section 2. Kuang-Fu Road, Hsinchu, 30013, \\
Taiwan (R.O.C.)\\
$^{3}$Department of Physics, National Chung Hsing University, 145 Xingda Rd., South Dist., Taichung 40227, Taiwan\\
$^{4}$Department of Space and Astronautical Science, Graduate University for Advanced Studies, SOKENDAI, Shonankokusaimura, Hayama, Miura District, \\ Kanagawa 240-0193, Japan\\
$^{5}$Institute of Space and Astronautical Science, Japan Aerospace Exploration Agency, 3-1-1 Yoshinodai, Chuo-ku, Sagamihara, Kanagawa 252-5210, Japan\\
$^{6}$Mullard Space Science Laboratory, University College London, Holmbury St. Mary, Dorking, Surrey, RH5 6NT, United Kingdom \\ 
$^{7}$National Centre for Nuclear Research, ul. Pasteura 7, 02-093 Warsaw, Poland\\
$^{8}$Astronomical Observatory of the Jagiellonian University, ul. Orla 171, 30-244 Cracow, Poland \\
$^{9}$Department of Physics and Astronomy, UCLA, 475 Portola Plaza, Los Angeles, CA 90095-1547, USA\\
$^{10}$Instituto de Astronom\'ia, Universidad Nacional Autónoma de México, AP 106, Ensenada 22860, Mexico\\
$^{11}$Department of Astronomy, Kyoto University, Kitashirakawa-Oiwake-cho, Sakyo-ku, Kyoto 606-8502, Japan\\
$^{12}$Academia Sinica Institute of Astronomy and Astrophysics, 11F of Astronomy-Mathematics Building, AS/NTU, No.1, Section 4, Roosevelt Road, Taipei 10617, Taiwan \\
$^{13}$Research Center for Space and Cosmic Evolution, Ehime University, 2-5 Bunkyo-cho, Matsuyama, Ehime 790-8577, Japan \\
$^{14}$Sabanc{\i} University, Faculty of Engineering and Natural Sciences, 34956, Istanbul, Turkey\\
$^{15}$Aix Marseille Univ. CNRS, CNES, LAM Marseille, France\\
$^{16}$Astronomy Program, Department of Physics and Astronomy, Seoul National University, 1 Gwanak-ro, Gwanak-gu, Seoul 08826, Republic of Korea\\
$^{17}$SNU Astronomy Research Center,
Seoul National University, 1 Gwanak-ro, Gwanak-gu, Seoul 08826, Republic of Korea\\
$^{18}$Korea Astronomy and Space Science Institute, 776 Daedeokdae-ro, Yuseong-gu, Daejeon 34055, Republic of Korea\\
$^{19}$Korea University of Science and Technology, 217 Gajeong-ro, Yuseong-gu, Daejeon 34113, Republic of Korea\\
$^{20}$Department of Earth Science Education, Kyungpook National University, 80 Daehak-ro, Buk-gu, Daegu 41566, Republic of Korea\\
$^{21}$RAL Space, STFC Rutherford Appleton Laboratory, Didcot, Oxon, OX11 0QX, UK\\
$^{22}$The Open University, Milton Keynes, MK7 6AA, UK\\
$^{23}$University of Oxford, Keble Rd, Oxford, OX1 3RH, UK\\
$^{24}$Department of Physics, National Tsing Hua University, No. 101, Section 2, Kuang-Fu Road, Hsinchu City 30013, Taiwan\\
}

\date{Accepted XXX. Received YYY; in original form ZZZ}

\pubyear{2021}

\begin{document}
\label{firstpage}
\pagerange{\pageref{firstpage}--\pageref{lastpage}}
\maketitle

\begin{abstract}
How does the environment affect active galactic nucleus (AGN) activity? We investigated this question in an extinction-free way, by selecting 1120 infrared galaxies in the \textit{AKARI} North Ecliptic Pole Wide field at redshift $z$ $\leq$ 1.2. A unique feature of the \textit{AKARI} satellite is its continuous 9-band infrared (IR) filter coverage, providing us with an unprecedentedly large sample of IR spectral energy distributions (SEDs) of galaxies. By taking advantage of this, for the first time, we explored the AGN activity derived from SED modelling as a function of redshift, luminosity, and environment. We quantified AGN activity in two ways: AGN contribution fraction (ratio of AGN luminosity to the total IR luminosity), and AGN number fraction (ratio of number of AGNs to the total galaxy sample). We found that galaxy environment (normalised local density) does not greatly affect either definitions of AGN activity of our IRG/LIRG samples (log ${\rm L}_{\rm TIR}$ $\leq$ 12).
However, we found a different behavior for ULIRGs (log ${\rm L}_{\rm TIR}$ $>$ 12). At our highest redshift bin (0.7 $\lesssim$ z $\lesssim$ 1.2), AGN activity increases with denser environments, but at the intermediate redshift bin (0.3 $\lesssim$ z $\lesssim$ 0.7), the opposite is observed. These results may hint at a different physical mechanism for ULIRGs. The trends are not statistically significant (p $\geq$ 0.060 at the intermediate redshift bin, and p $\geq$ 0.139 at the highest redshift bin). 
%Our results are more favorable towards the lack of local environmental influence on MIR AGN activity, although ULIRGs may exhibit a slightly different trend. 
Possible different behavior of ULIRGs is a key direction to explore further with future space missions (e.g., \textit{JWST}, \textit{Euclid}, \textit{SPHEREx}).
%may help resolve this problem by increasing the sample size.}
\end{abstract}


%%%%%%%%%%%%%%%%%%%%%%%%%%%%%%%%%%%%%%%%%%%%%%%%%%

%%%%%%%%%%%%%%%%% BODY OF PAPER %%%%%%%%%%%%%%%%%%

\section{Introduction}

An active galactic nucleus (AGN) is defined as a highly-luminous, compact object that radiates energy due to the conversion of gravitational energy to radiation via accretion of materials by a supermassive black hole (SMBH) which is believed to reside in the centre of most galaxies \citep[e.g.,][]{Caglar2020, Padovani2017, Rees1984, Ferrarese2005}. The active accretion of matter by SMBHs are most likely a result of increased matter density in their immediate environment \citep{KeelOswalt2013}. There are tight correlations between formation/evolution of SMBHs and their host galaxies. Some examples are the relation between the black hole mass and the host galaxy's stellar velocity dispersion ($M_{\rm BH}$ $\textrm{--}$ $\sigma_*$) \citep[e.g.,][]{Caglar2020} and the relation between the mass of the black hole and bulge mass of the host galaxy \citep[e.g.,][]{Marconi2003, McLure2002, Erwin2004}. These make studying AGNs crucial for understanding galaxy evolution.

Another area of interest for galaxy evolution is galaxy environmental effects. It is widely believed that cluster galaxies (galaxies that reside in clusters) have different properties compared to unclustered field galaxies due to environmental effects such as gravitational interactions with other galaxies and hydrodynamical effects like ram pressure stripping that are prevalent in cluster environments \citep{KeelOswalt2013, Boselli2006, Park2009}. Many studies have shown suggestive proof that galaxy evolution is closely affected by the environment. The environment influences many galaxy properties, such as star formation/colour \citep[e.g.,][]{Lemaux2019, Tomczak2019, Old2020}, morphology \citep[e.g.,][]{Dressler1980, Goto2003, Hwang2009, Kawinwanichakij2017, Sazonova2020}, and stellar mass \citep[e.g.,][]{Peng2010, Peng2012, Tomczak2017, vanderBurg2020}. Galaxy evolution is currently described by the hierarchical structure formation scenario of the Lambda Cold Dark Matter ($\Lambda$CDM) model, which is the canonical cosmological model of the Universe. In general, the structures observed in the Universe at present were formed by having smaller structures merged together to form larger structures \citep[e.g.,][]{Kauffmann2004}. In this case, galaxies are formed by having matter collapse into their own gravitational pull, and then the resulting galaxies further collapse by the gravity of the dark matter halos in a much larger scale hosting the galaxies to form galaxy clusters.

%For instance, the well known morphology-density relation indicates a decreasing number of spiral galaxies and an increasing number of elliptical galaxies with increasing local galaxy density \citep{Dressler1980}. On the other hand, other studies found a suggestive proof that galaxy evolution is closely affected by environment as it influences star formation, galaxy morphology, and stellar mass \citep[e.g.,][]{Alberts2016}. 

In this regard, the effects of environment in AGN activity is a critical topic of research. However, this connection is not well understood yet. The confusion among many studies focusing on this topic is mainly due to the difference in the definition of "environment" and how the samples are selected. As a consequence, miscellaneous conclusions result in no clear relationship between AGN activity and environment. 

For instance, in terms of local galaxy density, optically detected galaxies selected from the Sloan Digital Sky Survey (SDSS) within the redshift range 0.03 < $z$ < 0.1 and magnitude range 14.5 < $r$ < 17.77 have shown that the number of optical AGNs at fixed stellar mass is lower and star formation activity is weaker in denser environments (environments with higher local galaxy density), indicating the suppression of cold gas in denser environments \citep{Sabater2013}. Another study about optically detected SDSS galaxies selected in the redshift range 0.05 $\leq$ $z$ $\leq$ 0.95 and with brightness M($r^*$) $\geq$ $-20.0$ mag found that the fraction of star-forming galaxies (SFGs) decreases with density, but the AGN fraction (fraction of galaxies with AGNs) remains constant with local galaxy density, indicating the absence of relationship between SF and AGN activity \citep{Miller2003}. These results by \cite{Miller2003} are comparable with a previous study conducted with optically detected galaxies selected from the 15R-North galaxy redshift survey, a uniform spectroscopic survey (S/N ~ 10) with 3650 - 7400 Å spectral coverage and median redshift of $z$ = 0.05 showing that AGN fraction is insensitive to the surrounding galaxy density \citep{Carter2001}. When it comes to radio AGNs, radio galaxies with infrared (IR) detection in the redshift range 0.55 $\leq$ $z$ $\leq$ 1.30 drawn from the Observations of Redshift Evolution in Large-Scale Environments (ORELSE) survey show preference in various local density environments based on their AGN contribution fraction (ratio of AGN luminosity to the total IR luminosity) \citep{Shen2020}. In their work, sources with higher AGN contribution fraction show no local environmental preference, while sources with almost no AGN contribution prefer locally denser environments. This is contradictory to \cite{Sabater2013} who investigated radio AGNs detected in SDSS at 0.03 < $z$ < 0.1. They found out that radio AGN (nuclear) activity is enhanced at higher densities. X-ray AGNs also show various results. For example, X-ray AGNs detected in the Cosmic Evolution Survey (COSMOS) at 1.4 $\leq$ $z$ $\leq$ 2.5 show that at smaller scales (< 100 kpc), unobscured X-ray AGNs have more local neighbouring galaxies compared to obscured X-ray AGNs \citep{Bornancini2020}. This is opposite to the behavior of X-ray AGNs in the Ultimate XMM Extragalactic Survey (XXL) – South Field at 0.2 < $z$ < 1.0, wherein there are no significant differences between the local environments of obscured and unobscured X-ray AGNs \citep{Melnyk2018}.

%For instance, galaxies selected from the Sloan Digital Sky Survey (SDSS) within the redshift range 0.03 < $z$ < 0.1 and magnitude range 14.5 < $r$ < 17.77 have shown that the number of optical AGNs at fixed stellar mass is lower and star formation activity is weaker in denser environments (environments with higher local galaxy density), indicating the suppression of cold gas in denser environments \citep{Sabater2013}. Another study about SDSS galaxies selected in the redshift range 0.05 $\leq$ $z$ $\leq$ 0.95 and with brightness M($r^*$) $\geq$ $-20.0$ mag found that the fraction of star-forming galaxies (SFGs) decreases with density, but the AGN fraction (fraction of galaxies with AGNs) remains constant with local galaxy density, indicating the absence of relationship between SF and AGN activity \citep{Miller2003}. These results by \cite{Miller2003} are comparable with a previous study conducted with galaxies selected from the 15R-North galaxy redshift survey, a uniform spectroscopic survey (S/N \textasciitilde{}10) with 3650 - 7400 $\si{\angstrom}$ spectral coverage and median redshift of $z$ = 0.05 showing that AGN fraction is insensitive to the surrounding galaxy density \citep{Carter2001}.

Differences in the AGN activity in cluster and field environments are also inconclusive. \cite{Lopes2017} investigated AGNs from cluster members from the Northern Sky Optical Cluster Survey (NoSOCS) and field galaxies from SDSS at $z$ $\leq$ 0.1. They found that for massive host galaxies (log $M_*$ > 10.6), AGN number fraction (ratio of number of AGNs to the number of total galaxies) is higher in field environments. Looking into radio AGNs, galaxies selected from ORELSE survey within 0.55 < $z$ < 1.30 show that radio AGNs tend to reside within cluster cores and locally overdense environments \citep{Shen2017}. Radio AGNs selected from SDSS between 0.03 < $z$ < 0.3 were also shown to prefer cluster cores compared to field regions \citep{Best2005}. However, \cite{Magliocchetti2018a} showed that the AGN fraction of radio AGNs with mid-infrared (MIR) detection at $z$ $\leq$ 1.2 selected by the Very Large Array - Cosmic Evolution Survey (VLA-COSMOS) are smaller within clusters compared to field environments. For X-ray selected AGNs, previous studies have found that the fraction of galaxies with X-ray AGNs are smaller within clusters than in fields. This is the case for X-ray selected clusters from \textit{ROSAT} All Sky Survey at $z$ < 0.5 and their immediate surrounding field regions \citep{MishraDai2019}. \cite{Georgakakis2008}, on the other hand, showed the opposite trend: X-ray AGNs at 0.7 < $z$ < 1.4 prefer group environments rather than isolated environments. They speculated that this is due to AGNs being hosted by red luminous galaxies which prefer denser environments. However, a study using a spectroscopically complete sample of 35 AGNs in 8 galaxy clusters between 0.06 < $z$ < 0.31 selected by \textit{Chandra} X-Ray Observatory showed that the fraction of galaxies with AGNs is comparable for both cluster and field galaxies \citep{Martini2007}. This same conclusion was also found by \cite{Haggard2010}, where they investigated the clusters at $z$ = 0.7 in the \textit{Chandra} Multiwavelength Project (ChaMP), at least for samples with similar redshift and absolute magnitude ranges.

The distance from the cluster center (or clustercentric distance) can also probe galaxy environment. However, its relationship with AGN activity is still unclear. For optical AGNs, \cite{Best2007} used galaxies selected from SDSS at 0.02 < $z$ < 0.1 and showed that the number fraction of galaxies with optical-emission line AGN activity decreases within the virial radius in galaxy groups and clusters. In contrast, optically selected galaxies from SDSS at a similar redshift range show that the number fraction of SFGs with powerful optical AGNs is independent of clustercentric distance \citep{vonderLinden2010}. \cite{Best2007} also studied radio AGNs and showed that only at small clustercentric distances (within one-fifth of the virial radius) do cluster galaxies show enhanced radio-AGN activity. X-ray AGN number fraction, on the other hand, is shown to increase with increasing clustercentric distance. This is true for AGNs in X-ray selected clusters from \textit{ROSAT} All Sky Survey at $z$ < 0.5 \citep{MishraDai2019} and X-ray AGNs in clusters identified in the Boötes field of the National Optical Astronomy Observatory (NOAO) Deep Wide-Field Survey \citep{Galametz2009}. However, AGNs selected in 6 clusters at $z$ < 0.08 by Chandra X-Ray Observatory showed that AGN number fraction does not change with clustercentric distance \citep{Sivakoff2008}.

%As for the clustercentric distance, several studies have shown that the AGN number fraction increases with increasing clustercentric distance. \textcolor{red}{This is true for AGNs in X-ray selected clusters from \textit{ROSAT} All Sky Survey at z < 0.5 \citep{MishraDai2019} and X-ray AGNs in clusters identified in the Boötes field of the National Optical Astronomy Observatory (NOAO) Deep Wide-Field Survey \citep{Galametz2009}.} This was the case for X-ray selected clusters from the \textit{ROSAT} and their immediate surrounding field regions below $z$ < 0.5 \citep{MishraDai2019}. However, studies about star-forming galaxies with powerful optical AGN selected in a sample of SDSS clusters at $z$ < 0.1 \citep{vonderLinden2010}, and AGNs selected in 6 clusters at $z$ < 0.08 by \textit{Chandra X-Ray Observatory} \citep{Sivakoff2008} showed that AGN number fraction does not change with clustercentric distance. Different correlations are found for different galaxy morphologies \citep[e.g.][]{Hwang2012a}, and different types of AGNs \citep[e.g.][]{MonteroDorta2009}.

Dark matter halo masses derived from halo occupation models can also probe galaxy environments. By investigating optical and radio AGNs from SDSS at 0.01 < z < 0.3, \cite{Mandelbaum2009} showed that optical AGNs and galaxies without AGNs with similar stellar masses have similar dark matter halo masses. Radio AGNs inhabit much more massive dark matter haloes compared to their non-AGN counterparts with similar stellar masses. On the other hand, direct measurement of mean halo occupation distribution (HOD) suggests that the number of X-ray AGNs do not grow as quickly as the halo mass. This is at least true to X-ray AGNs with dark matter halo masses within the range of log ${\rm M}_{\rm h}$ [${\rm M}_\odot$] = 13-14.5 that are detected in XMM and C-COSMOS ($z$ $\leq$ 1), and \textit{ROSAT} All-Sky Survey (0.16 < $z$ < 0.36) \citep{Miyaji2011, Allevato2012}. 

%Finally, direct measurement of mean halo occupation distribution (HOD) also suggests that the number of X-ray AGNs do not grow as quickly as the halo mass. This is at least true to X-ray AGNs with dark matter halo masses within the range of log $M_h$[$M_\odot$] = 13-14.5 that are detected in XMM and C-COSMOS ($z$ $\leq$ 1), and \textit{ROSAT} All-Sky Survey (0.16 < $z$ < 0.36) \citep{Miyaji2011, Allevato2012}. All of these results with different environmental parameters are examples indicating that the dependence of AGN fraction on environment differs for AGNs selected in different wavelengths and redshifts, as well as the environmental parameter(s) being used.

All these results with different environmental parameters are examples indicating that the dependence of AGN fraction on environment differs for AGNs selected in different wavelengths and redshifts, as well as the environmental parameter(s) being used. To further investigate the relationship between AGN activity and environment, it is best to also observe the hidden AGN activity that is not detected in other spectral regions \citep{Hickox2018}. AGNs can be observed in different spectral wavelengths, namely radio, IR, optical, and X-ray. However, heavily obscured populations of Type II (obscured) AGNs cannot be detected in optical and soft X-ray surveys 
due to extinction/absorption by a large amount of dust and gas particles. 
In addition, only \textasciitilde{}10\% of AGNs are radio-loud and are detected in radio wavelengths, and current X-ray telescopes are not sensitive enough to detect them \citep{Chiang2019}. It is therefore crucial to use infrared, particularly MIR, since MIR provides the distinctive diagnostic for identifying the hidden AGN activity behind the dusty torus, which re-emits AGN radiation into thermal emission \citep{Gonzalez-Martin2019a, Gonzalez-Martin2019b}. %since MIR surveys do not also suffer from extinction.
However, SFGs are also detected in the MIR region \citep{Kim2019}, which makes identifying AGNs more challenging. The spectra of SFGs differ from AGNs since the AGN spectrum resembles a red power-law spectrum ($f_\nu \propto \nu^{-\alpha}$ where $\alpha$ is typically less than -0.5) \citep{AlonsoHerrero2003}. This dilutes the polycyclic aromatic hydrocarbon (PAH) emission features in the MIR which are usually found in SFGs \citep{Jensen2017, Kim2019, Ohyama2018}. Looking at the MIR spectral energy distributions (SEDs) of the sources can therefore help us identify AGNs from SFGs \citep[e.g.,][]{Laurent2000, Huang2017, Wang2020, Toba2020a}, justifying the utilisation of MIR surveys for studying AGNs.

The Wide-field  Infrared  Survey  Explorer (\textit{WISE}) \citep{Hwang2012b}, \textit{Spitzer} \citep{Toba2015, Toba2016}, and \textit{AKARI} \citep{Murakami2007} are some of the infrared telescopes that can be utilised for this task. However, both \textit{WISE} and \textit{Spitzer} have limited number of available filters and gaps/discontinuities in the bandwidth of the filters: \textit{WISE} has only 4 filters with reference wavelengths of 3.4, 4.6, 12.0 and 22.0 $\mu$m, while \textit{Spitzer} has only 5 filters with reference wavelengths of 3.6, 4.5, 5.8, 8.0 and 24.0 $\mu$m (considering only MIR). A gap in their filter system in the MIR wavelength range (between $12- 22\,\mu$m and $8-24\,\mu$m in \textit{WISE} and \textit{Spitzer}, respectively) can also cause difficulties in distinguishing SFGs and AGNs especially if the features of a source lie within these wavelength ranges. The \textit{AKARI} telescope's Infrared Camera (IRC; \citealt{Onaka2007}), on the other hand, does not suffer from this problem by virtue of its continuous 9-band filter coverage ranging from near-IR to far-IR (N2, N3, N4, S7, S9W, S11, L15, L18W, and L24, with reference wavelengths of  2.4, 3.2, 4.1, 7.0, 9.0, 11.0, 15.0, 18.0, and 24.0 , respectively). This makes \textit{AKARI} efficient in finding obscured AGNs. Furthermore, the availability of multiwavelength data of our sources detected by \textit{AKARI} allows us to reproduce as large as one order of magnitude more SEDs compared to previous infrared spectrographs (e.g., \textit{Spitzer}/InfraRed Spectrograph (IRS)) which can only achieve \textasciitilde{}100 galaxies at most \citep{Wang2011}. Many studies have taken advantage of \textit{AKARI} \citep[e.g.,][]{Huang2017, Chiang2019, Wang2020, Miyaji2019, Toba2020b, KilerciEser2020}, and this work will also make use of it for the same aforementioned purpose.

This study aims to investigate the AGN activity-environment relation for galaxies in the \textit{AKARI} North Ecliptic Pole Wide (NEPW) field. This work is organised as follows: Sec.~\ref{sec:data} contains the description of the sample selection criteria and required calculations for this study, Sec.~\ref{sec:results} shows the results of our work, Sec.~\ref{sec:discussion} is dedicated to discussion, and finally, Sec.~\ref{sec:conclusion} presents the conclusion of this study. We assumed a flat cosmology with ${\rm H}_{\rm 0}$ = 70.4 km ${\rm s}^{-1}$ ${\rm Mpc}^{-1}$, $\Omega_\Lambda$ = 0.728, and $\Omega_{\rm M}$ = 0.272 \citep{Komatsu2011}.

\section{Data and Analysis}
\label{sec:data}

\subsection{Sample Selection}
\label{sec:sample_selection}

\begin{figure*}
	\includegraphics[width=2\columnwidth]{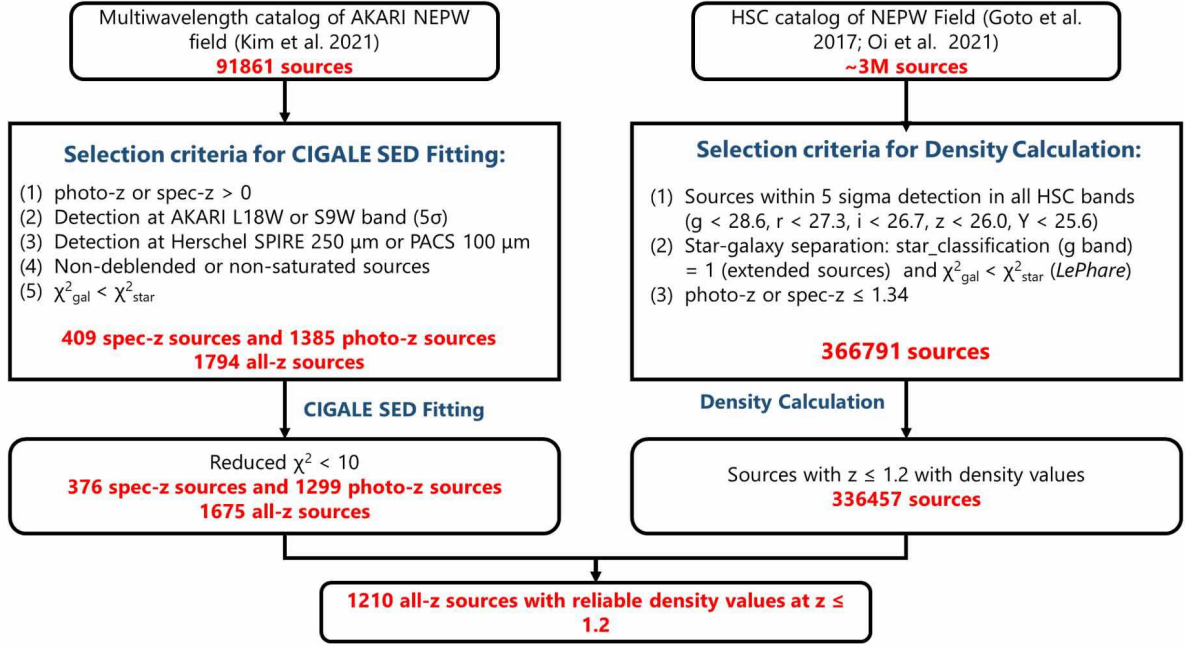}
    \caption{Sample selection flowchart for this study. The left side of the flowchart summarises the MIR-FIR source selection for investigating obscuration-free AGN activity. The right side, on the other hand, summarises the source selection in the optical wavelengths for measuring galaxy density.}
    \label{fig:sample_selection}
\end{figure*}

There are two calculations that require different sets of sample selection criteria: SED fitting for constraining the galaxy properties of our sources, and density calculation for constraining the galaxy environment of our sources. The former is the sample for the main analyses. The latter sample is more abundant, and thus more reliable for the density calculation.
Fig.~\ref{fig:sample_selection} shows a summary of our selection criteria. All sources for each calculation are situated within the NEPW field. In this section, further explanation about the sample selection is presented. 

On the left side of Fig~\ref{fig:sample_selection}, the selection criteria for constraining the galaxy properties, including hidden AGN activity, of our sources are shown. All of these sources are selected from a multiwavelength catalogue \citep{Kim2021} in the NEPW field survey \citep{Matsuhara2006, Lee2007, Kim2012} of \textit{AKARI} which contains 91,861 sources. The NEPW survey carried out an observation of 5.4 deg$^{2}$ area at R.A.=18${h}$00${m}$00${s}$, Dec.=+66${d}$33${m}$38${s}$ by \textit{AKARI} Infrared Camera (IRC). The \textit{AKARI} NEPW field catalogue was created by cross-matching data from \textit{AKARI} and from Subaru Hyper Suprime-Cam (HSC; \cite{Miyazaki2018}), enabling the identification of IR sources from \textit{AKARI} using deep optical data from HSC, with additional counterpart data from optical to far-infrared (FIR) wavelengths. The multiwavelength analysis is crucial for achieving better constraint of the sources' galaxy properties. More details about the band-merging can be read from \cite{Kim2021}.

This work used similar selection criteria as those presented by \cite{Wang2020}. To summarise, in their work, the sources must be detected at the \textit{AKARI} \textit{L18W} band (18.0 $\mu$m) and in \textit{Herschel}/PACS Green (100 $\mu$m) or \textit{Herschel}/SPIRE PSW (250 $\mu$m) band to assure proper constraint of the templates used for SED fitting. The detection requirement in \textit{Herschel}/PACS Green band or \textit{Herschel}/SPIRE PSW band is implemented to make sure that there is at least one observation in the FIR region. The requirement of AKARI \textit{L18W} band detection, whose bandwidth is in the longer part of the MIR band (MIR-L), is driven by its limited sensitivity compared to the other MIR bands in \textit{AKARI}. To increase the sample, we also add another criterion which is the detection in \textit{AKARI} \textit{S9W} band. The number of detections in this band is the largest among the shorter part of the MIR bands (MIR-S). Moreover, \textit{S9W} has a wide filter width which covers the wavelength ranges of \textit{S7} and \textit{S11} bands \citep{Kim2012}, making it a viable choice for making sure that at least one among the shorter and longer parts of the MIR band is available for all sources. The availability of MIR and FIR data is important in making sure that the SED fitting is robust via the energy balance principle (see Sec.~\ref{sec:CIGALE} for an explanation). 

The photometric redshifts (photo-$z$) of all the sources in the \textit{AKARI} NEPW field without spectroscopic redshifts (spec-$z$) were provided by \cite{Ho2021}. \cite{Ho2021} performed the photo-$z$ estimation with Photometric Analysis for Redshift Estimate \citep[$Le$ $Phare$;][]{Arnouts1999, Ilbert2006}, an SED template-fitting code. In this study, we used spec-$z$ as the redshift values of the sources if they were available; otherwise, we used their estimated photo-$z$. We selected the galaxies in our sample using the $\chi^2$ value provided by $Le$ $Phare$: if $\chi^2_{\rm gal} < \chi^2_{\rm star}$ (where $\chi^2_{\rm gal}$ and $\chi^2_{\rm star}$ are the minimum $\chi^2$ values calculated using the galaxy and stellar templates, respectively), the object is most likely a galaxy. We also removed sources with erroneous redshift values (negative photo-$z$ and/or spec-$z$) and deblended or saturated sources as suggested by \cite{Ho2021}.

For \texttt{CIGALE} SED fitting (see Sec. \ref{sec:CIGALE}), the reduced chi-square ($\chi^2$) of the best-fit SED of each source must be less than 10 to make sure that the estimated properties of the sources are reliable. The value of 10 was chosen as the threshold for the reduced $\chi^2$ value as a result of our visual checks in our SED fitting results. In addition, it was also used by \cite{Wang2020} because of the fact that there are at most 36 photometric points used to estimate the SED of the sources and they are very likely to be correlated, causing the $\chi^2$ value to not be well-estimated \citep{Wang2020}. We also investigated the reliability of our constrained galaxy properties (i.e., AGN contribution fraction and star formation rate) by running \texttt{CIGALE}'s mock analysis. The results of our mock analysis is shown in Appendix \ref{sec:appendixA}. Our mock analysis shows that we were able to constrain the AGN contribution fractions and star formation rates of our sources with enough reliability, as the estimated (mock) quantities reasonably agree with the exact (true) quantities.

Before the reduced $\chi^2$ cut, we initially acquired 1794 all-$z$ sources which are composed of 409 spec-$z$ sources and 1385 photo-$z$ sources. Only 119/1794 (6.6\%) sources were removed with this criteria, and we eventually acquired 1675 all-$z$ sources which are composed of 376 spec-$z$ sources and 1299 photo-$z$ sources. The median reduced $\chi^2$ of our all-z sample after the reduced chi-square cut is 3.63.

On the right side of Fig~\ref{fig:sample_selection}, the selection criteria for constraining the environment of our sources is shown. Density was calculated using sources in the AKARI NEPW field detected within 5$\sigma$ detection limits (with magnitude errors less than 0.2) in all bands of HSC. Because it has the largest field-of-view (FoV) among optical cameras on an 8m telescope (1.5 deg in diameter), HSC was able to carry out an optical survey of the AKARI NEPW field with just 4 FoVs \citep{Goto2017, Oi2021}. The 5-$\sigma$ detection limits of the HSC bands $g$, $r$, $i$, $z$, and $Y$ are 28.6, 27.3, 26.7, 26.0, and 25.6 (in AB mag), respectively. There are a total of $\sim$3 million (3M) optically-detected sources in the AKARI NEPW field.

As mentioned earlier, only 2026 sources have spec-$z$, and so the rest must rely on photo-$z$ to have their density calculated. The photo-$z$ of the HSC-detected sources without spec-$z$ (that are not included in the \textit{AKARI} NEPW field catalog) were calculated using $Le$ $Phare$. To improve the photo-$z$ of the sources, the following data were also added for estimating photo-$z$: \textit{u}imaging data of Canada-France-Hawaii Telescope (CFHT; \citealt{Iye2003}) obtained and reduced by \cite{Huang2020}, and deep \textit{Spitzer} Infrared Array Camera (IRAC) (3.6 $\mu$m- and 4.5$\mu$m-band) data from \cite{Nayyeri2018}. By comparing with sources with spec-$z$, the photo-$z$ of the HSC-detected sources reached a photo-$z$ dispersion of 0.064 and an outlier fraction ($\eta$, defined as the number ratio of outliers with |$\Delta$z|/(1 + $z_s$)> 0.15 to the total number of sources) of 9.1$\%$ for sources with $z$ $<$ 1.5. \cite{Lai2016} showed that a photo-$z$ uncertainty of 0.06 is still acceptable for seeing the dependence of red fraction on galaxy environment for sources at redshift $z$ \textasciitilde{} 0.8. Although red fraction is not the focus of this paper, a photo-$z$ uncertainty of 0.064 is good enough for studying the effect of galaxy environment on AGN activity.

In addition, galaxy density should be calculated with galaxies only, so star-galaxy separation was necessary. Galaxies were selected based on two star-galaxy separation criteria: the HSC pipeline parameter \texttt{base$\_$ClassificationExtendedness$\_$value} (to remove extended sources), and the chi-square ($\chi^2$) values of the sources using both the galaxy templates and stellar SED templates provided by $Le$ $Phare$ (similar to that from SED Fitting selection criteria). 

We would like to limit our study up to $z$ $\leq$ 1.2 since the photo-$z$ performance becomes worse at higher redshift. However, we did not want to lose sources at $z$ $>$ 1.2 when calculating density of sources at $z$ $=$ 1.2. At $z$ = 1.2, the redshift bin size (further discussed in Sec.~\ref{sec:environment}) is 0.14, so we restricted our density calculation up to $z$ = 1.2 + 0.14 = 1.34. This criteria gave us 366791 sources for density calculation. Finally after density calculation, we have 336457 sources within z $\leq$ 1.2 with density values.

\subsection{Estimation of Galaxy Physical Properties}
\label{sec:CIGALE}

\texttt{CIGALE}\footnote{\url{https://CIGALE.lam.fr}} \citep{Boquien2019}, short for Code Investigating GALaxy Emission, is an SED fitting and modelling code that can handle many parameters such as dust thermal emission, star formation history (SFH), single stellar population (SSP), attenuation law, and AGN emission. We follow \cite{Wang2020}'s approach in using \texttt{CIGALE} SED fitting to select AGNs in the \textit{AKARI} NEPW field and analyse the dependence of the sources' AGN activity on redshift and luminosity. We used \texttt{CIGALE} version 2018.0 to model the optical to far-IR emission of each source. \texttt{CIGALE} is mainly centred on the energy balance principle: the energy absorbed by dust in the UV and optical wavelengths is re-emitted in the mid- and far-IR spectral regions \citep[e.g.][]{Burgarella2005, Noll2009, Boquien2019}. In this regard, it is important to have sufficient observations in mid- and far-IR wavelengths to accurately measure the absorbed and re-emitted energy. The physical modelling by \texttt{CIGALE} is used to fit the stellar, AGN, and SF components of each source's SED, which enables us to get more information about the sources' properties. 

\begin{figure}
	\includegraphics[width=\columnwidth]{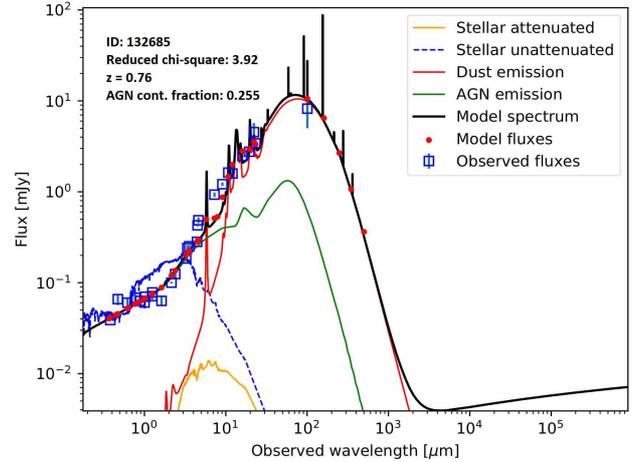}
    \caption{An example SED fitting result from \texttt{CIGALE}. The AKARI ID, reduced chi-square, redshift, and AGN contribution fraction of the source are displayed on the upper left corner.}
    \label{fig:sample_sed}
\end{figure}

This work used the same modules and parameters as \cite{Wang2020}. We used the delayed SFH module with optional exponential burst and parameterised age of main stellar population in the galaxy. The e-folding times of the main stellar population ($\tau_{\rm main}$) and the late starburst population ($\tau_{\rm burst}$) were also fixed. In addition, we also utilised the stellar templates from \cite{Bruzual2003} assuming \cite{Salpeter1955}'s initial mass function. The standard default nebular emission model from \cite{Inoue2011} was also used. The dust attenuation, on the other hand, was modelled using \cite{Charlot2000}'s attenuation law model. This model assumes that there are two power-law attenuation curves for the birth cloud and the interstellar medium (ISM) that differ on their default slopes. We made the slopes of these curves flexible, turning the attenuation law similar to a double power law \citep[e.g.][]{LoFaro2017, Buat2019}. The V-band attenuation in the ISM ($A_{\rm V}^{\rm ISM}$) was separately parameterised but the ratio of the old stars' V-band attenuation to the young stars' V-band attenuation ($\mu$) was fixed to $\mu$ = 0.44 \citep{Malek2018}. \cite{Draine2013}'s dust emission model was utilised for our SED fitting to model the reprocessed IR dust emission absorbed from UV/optical stellar emission. Lastly, we used the AGN emission model module by \citet{Fritz2006} which considers three main components using a radiative transfer model: (1) the primary source of radiation located inside the torus, (2) the dust-scattered radiation emission, and (3) the thermal dust emission \citep{Boquien2019}. Table~\ref{tab:cigale} shows the parameter settings that we used in our \texttt{CIGALE} SED fitting. Fig.~\ref{fig:sample_sed} shows a sample SED fitting result with \texttt{CIGALE}.

\begin{table*}
	\centering
	\caption{List of modules and parameter settings for our \texttt{CIGALE} SED fitting.}
	\label{tab:cigale}
	\begin{tabular}{cc}
		\hline
		\textbf{Parameters} & \textbf{Value} \\
		\hline
		\multicolumn{2}{c}{\textit{Delayed SFH with optional exponential burst}} \\
		\hline
		e-folding time of the main stellar population [$10^6$ yr] & 5000.0 \\
		Age of the galaxy's main stellar population [$10^6$ yr] & 1000.0, 5000.0, 10000.0 \\
		e-folding time of the late starburst population [$10^6$ yr] & 20000.0 \\
		Age of the late burst [$10^6$ yr] & 20.0 \\
		Mass fraction of the late burst population & 0.00, 0.01 \\
		\makecell{Multiplicative factor controlling the amplitude of SFR if \\ normalisation is false} & 1.0 \\
		\hline
		\multicolumn{2}{c}{\textit{SSP} \citep{Bruzual2003}} \\
		\hline
		Initial mass function & \cite{Salpeter1955} \\
		Metallicity & 0.02 \\
		Age of separation between the young and old star populations & 10.0 \\
		\hline
		\multicolumn{2}{c}{\textit{Dust attenuation} \citep{Charlot2000}} \\
		\hline
		Logarithm of the V-band attenuation in the ISM & -2.0, -1.7, -1.4, -1.1, -0.8, -0.5, -0.2, 0.1, 0.4, 0.7, 1.0 \\
		Ratio of V-band attenuation from old and young stars & 0.44 \\
		Power-law slope of the attenuation in the ISM & -0.9, -0.7, -0.5 \\
		Power-law slope of the attenuation in the birth cloud & -1.3, -1.0, -0.7 \\
		\hline
		\multicolumn{2}{c}{\textit{Dust emission} \citep{Draine2013}} \\
		\hline
		Mass fraction of PAH & 0.47, 1.77, 2.50, 5.26 \\
		Minimum radiation field (${\rm U}_{\rm min}$) & 0.1, 1.0, 10.0, 50.0 \\
		Power-law slope $\alpha$ ($\frac{{\rm dU}}{{\rm dM}} \propto U^\alpha$) & 1.0, 1.5, 2.0, 2.5, 3.0 \\
		Fraction illuminated from ${\rm U}_{\rm min}$ to ${\rm U}_{\rm max}$ & 1.0 \\
		\hline
		\multicolumn{2}{c}{\textit{AGN emission} \citep{Fritz2006}} \\
		\hline
		Ratio of the maximum and minimum torus radii & 60.0 \\
		Optical depth at 9.7 $\mu$m & 0.3, 6.0 \\
		\makecell{Value of $\beta$ in gas density gradient along the \\ radial and polar distance coordinates (Eqn. 3 in \cite{Fritz2006})} & -0.5 \\
		\makecell{Value of $\gamma$ in gas density gradient along the \\ radial and polar distance coordinates (Eqn. 3 in \cite{Fritz2006})} & 4.0 \\
		Opening angle of the torus & 100.0 \\
		Angle between equatorial axis and line of sight & 0.001, 60.100, 89.900 \\
		AGN contribution fraction (${\rm frac}_{\rm AGN}$) & \makecell{0.0, 0.025, 0.05, 0.075, 0.1, 0.125, 0.15, 0.175, 0.2, 0.225,0.25, \\ 0.275, 0.3, 0.325, 0.35, 0.375, 0.4, 0.425, 0.45, \\ 0.475, 0.5, 0.525, 0.55, 0.575, 0.6, 0.625, 0.65, 0.675, 0.7} \\
		\hline
	\end{tabular}
\end{table*}

\begin{table*}
	\centering
	\caption{Summary of the 36 bands used in this study.}
	\label{tab:bands}
	\begin{tabular}{cccccc}
	\hline
		\textbf{\makecell{Wavelength \\ Range}} & \textbf{Band} & \textbf{Instrument} & \textbf{\makecell{Wavelength \\ ($\mu$m)}} & \textbf{\makecell{Depth \\ (AB Mag)}} & \textbf{Reference} \\
	\hline
		\multirow{10}{*}{\makecell{UV to \\ Optical}} 
		& $g$ & $Subaru$/HSC & 0.47 & 28.6 & \cite{Oi2021} \\
		& $r$ & $Subaru$/HSC & 0.61 & 27.3 & \cite{Oi2021} \\
		& $i$ & $Subaru$/HSC & 0.77 & 26.7 & \cite{Oi2021} \\
		& $z$ & $Subaru$/HSC & 0.89 & 26.0 & \cite{Oi2021} \\
		& $Y$ & $Subaru$/HSC & 0.97 & 25.6 & \cite{Oi2021} \\
		& $u*$ & CFHT/MegaCam & 0.39 & 26, 24.6 & \cite{Hwang2007, Oi2014} \\
		& $u$ & CFHT/MegaPrime & 0.38 & 25.4 & \cite{Huang2020} \\
		& $B$ & Maidanak/SNUCAM & 0.44 & 23.4 & \cite{Jeon2010} \\
		& $R$ & Maidanak/SNUCAM & 0.61 & 23.1 & \cite{Jeon2010} \\
		& $I$ & Maidanak/SNUCAM & 0.85 & 22.3 & \cite{Jeon2010} \\
	\hline
	    \multirow{15}{*}{\makecell{Near to \\ Mid-IR}}
	    & $N2$ & $AKARI$/IRC & 2.3 & 20.9 & \cite{Kim2012} \\
	    & $N3$ & $AKARI$/IRC & 3.2 & 21.1 & \cite{Kim2012} \\
	    & $N4$ & $AKARI$/IRC & 4.1 & 21.1 & \cite{Kim2012} \\
	    & $S7$ & $AKARI$/IRC & 7.0 & 19.5 & \cite{Kim2012} \\
	    & $S9W$ & $AKARI$/IRC & 9.0 & 19.3 & \cite{Kim2012} \\
	    & $S11$ & $AKARI$/IRC & 11.0 & 19.0 & \cite{Kim2012} \\
	    & $L15$ & $AKARI$/IRC & 15.0 & 18.6 & \cite{Kim2012} \\
	    & $L18W$ & $AKARI$/IRC & 18.0 & 18.7 & \cite{Kim2012} \\
	    & $L24$ & $AKARI$/IRC & 24.0 & 17.8 & \cite{Kim2012} \\
	    & $H$ & KPNO/FLAMINGOS & 1.6 & 21.3 & \cite{Jeon2014} \\
	    & $J$ & KPNO/FLAMINGOS & 1.3 & 21.6 & \cite{Jeon2014} \\
	    & $W1$ & $WISE$/ALLWISE & 3.4 & 18.1 & \cite{Jarrett2011} \\
	    & $W2$ & $WISE$/ALLWISE & 4.6 & 17.2 & \cite{Jarrett2011} \\
	    & $W3$ & $WISE$/ALLWISE & 12.0 & 18.4 & \cite{Jarrett2011} \\
	    & $W4$ & $WISE$/ALLWISE & 22.0 & 16.1 & \cite{Jarrett2011} \\
	    \multirow{7}{*}{\makecell{Near to \\ Mid-IR}}
	    & $Y$ & WIRCam & 1.02 & 23.4 & \cite{Oi2014} \\
	    & $J$ & WIRCam & 1.25 & 23.0 & \cite{Oi2014} \\
	    & $Ks$ & WIRCam & 2.14 & 22.7 & \cite{Oi2014} \\
	    & IRAC1 & $Spitzer$ & 3.6 & 21.8 & \cite{Nayyeri2018} \\
	    & IRAC2 & $Spitzer$ & 4.5 & 22.4 & \cite{Nayyeri2018} \\
	    & IRAC3 & $Spitzer$ & 5.8 & 16.6 & \cite{Nayyeri2018} \\
	    & IRAC4 & $Spitzer$ & 8.0 & 15.4 & \cite{Nayyeri2018} \\
	\hline
	    \multirow{5}{*}{Far-IR}
	    & PSW & $Herschel$/SPIRE & 253.7 & 14.0 & \cite{Pearson2019} \\
	    & PMW & $Herschel$/SPIRE & 169.8 & 14.2 & \cite{Pearson2019} \\
	    & PLW & $Herschel$/SPIRE & 253.7 & 13.8 & \cite{Pearson2019} \\
	    & PACS Green & $Herschel$/SPIRE & 103.7 & 14.7 & \cite{Pearson2017} \\
	    & PACS Red & $Herschel$/PACS & 169.8 & 14.1 & \cite{Pearson2017} \\
	\hline
	\end{tabular}
\end{table*}

We also used the same bands as \cite{Wang2020} for SED fitting. We have all 9 IR bands from \textit{AKARI}/IRC \citep{Kim2012}, $u^{*}$-band from CFHT/MegaCam \citep{Hwang2007, Oi2014, Huang2020}, all 5 bands from HSC \citep{Oi2021}, $B$, $R$ and $I$ bands from Maidanak/Seoul National University 4K x 4K Camera (SNUCAM) \citep{Jeon2010}, $H$- and $J$-bands from Kitt Peak National Observatory (KPNO)/Florida Multi-object Imaging Near-IR Grism Observational Spectrometer (FLAMINGOS) \citep{Jeon2014}, $Y$, $J$, and $K$ bands from CFHT/Wide-field InfraRed Camera (WIRCam) \citep{Oi2014}, all 4 IR bands from \textit{WISE}/ALLWISE \citep{Jarrett2011}, all 4 bands from \textit{Spitzer}/IRAC \citep{Nayyeri2018}, Green and Red bands from \textit{Herschel}/PACS \citep{Pearson2019}, and all 3 FIR bands from \textit{Herschel}/SPIRE \cite{Pearson2017}. More details about their effective wavelengths, area, and detection limits/depth are listed in Table \ref{tab:bands}.

In this work, we defined two probes of AGN activity following \cite{Wang2020}: AGN contribution fraction and AGN number fraction. AGN contribution fraction (${\rm frac}_{\rm AGN}$) is related to the AGN luminosity in the 8-1000 $\mu$m range (${\rm L}_{\rm AGN}$) and total IR luminosity (${\rm L}_{\rm TIR}$): 
\begin{eqnarray}
    {\rm L}_{\rm AGN}= {\rm L}_{\rm TIR} \times {\rm frac}_{\rm AGN}.
	\label{eq:agncontributionfrac}
\end{eqnarray}

On the other hand, AGN number fraction (${\rm N}_{\rm AGN}/{\rm N}_{\rm tot}$) is the ratio between the number of AGNs, ${\rm N}_{\rm AGN}$, and total number of galaxies, $N_{\rm tot} = {\rm N}_{\rm SFG+AGN}$ within a certain bin (e.g. density, luminosity, redshift):
\begin{eqnarray} 
   {\rm N}_{\rm AGN}/{\rm N}_{\rm tot} = \frac{N_{\rm AGN}}{N_{\rm SFG+AGN}}
   \label{eq:agnnumberfrac}
\end{eqnarray}

\cite{Wang2020} focused on finding the relationship between AGN activity (${\rm frac}_{\rm AGN}$ and ${\rm N}_{\rm AGN}/{\rm N}_{\rm tot}$) and total IR luminosity/redshift for MIR sources in the \textit{AKARI} NEPW field. This was made possible by using \texttt{CIGALE} to constrain galaxy properties and select AGNs. They compared the number of sources with varying ${\rm frac}_{\rm AGN}$ with AGNs classified by spectroscopic surveys and \textit{Chandra} X-ray surveys in \cite{Shim2013} and \cite{Krumpe2015}, and AGNs classified using $Le$ $Phare$ \cite{Oi2021}. A similar study by \cite{Shen2020} also used \texttt{CIGALE} to select AGNs at 0.55 $\leq$ $z$ $\leq$ 1.30 in the Observations of Redshift Evolution in Large Scale Environments (ORELSE) survey, and they adopted the definition ${\rm frac}_{\rm AGN}$ $\geq$ 0.1 as their AGN selection criteria. However, in \cite{Wang2020}'s comparison of previously identified AGNs in the \textit{AKARI} NEPW field, they decided to select AGNs with ${\rm frac}_{\rm AGN}$ $\geq$ 0.2 because majority of the sources with ${\rm frac}_{\rm AGN}$ between 0.1 and 0.2 (69\%) are not classified as AGNs in both spectroscopic and X-ray classifications. Another key result \cite{Wang2020} found is that AGN activity increases with redshift and not with luminosity, inferring that we may find more AGNs at higher redshift. Using ${\rm frac}_{\rm AGN}$ $\geq$ 0.2 as our AGN selection criteria allowed us to select only 35/1210 AGNs. This can be attributed to our limited redshift range ($z$ $\leq$ 1.2). Therefore, for discussion purposes, we also adopted a less strict threshold (${\rm frac}_{\rm AGN}$ $\geq$ 0.15) for AGN selection that compromises the number of selected AGNs and accuracy based on \cite{Wang2020}'s previous comparison. In this study, we followed these two AGN selection criteria to also check whether our results change with varying threshold of AGN selection or not.

Note that AGN contribution fraction and AGN number fraction are different physical quantities. AGN contribution fraction is closely related to the AGN's strength relative to its host galaxy's luminosity, while AGN number fraction describes the number of AGN population with respect to a certain quantity (in this case, galaxy environment). One of them may exhibit different relationship with density compared to the other \citep{Wang2020}, and so looking into both quantities is crucial for a better understanding of AGN activity and their relationship with galaxy environment.

\subsection{Density Estimation}
\label{sec:environment}

Local density was first used by \cite{Dressler1980} to study galaxy morphology in clusters. It was also used by many other works due to its usefulness in studying galaxy environments in groups and clusters \citep[e.g.][]{Dressler1980, Lewis2002} and its capability to use redshift information of the sources to exclude foreground and background sources by limiting the involved galaxy neighbours to a specific velocity interval \citep{Cooper2005}. In addition, it provides a continuous (non-discrete) measurement of environment, which is advantageous compared to classifying galaxies into predetermined environment categories \citep{Cooper2005}. In this work, we used the 10th-nearest neighbourhood method to calculate the local galaxy density of our sample (see Sec.~\ref{sec:sample_selection} for the sample selection for density calculation) which is described by Equation~\ref{eq:localdensity}:
\begin{eqnarray}
    \Sigma_{10}=\frac{10}{{\theta_{10}}^2}
	\label{eq:localdensity}
\end{eqnarray}
where $\theta_{10}$ is the angular separation/projected distance to the 10th nearest neighbour (measured in Mpc). As for excluding foreground and background sources, we are limited by the very few sources with spectroscopic redshifts (spec-$z$). Only 2026 of the sources in the \textit{AKARI} NEPW field have spec-$z$ provided by \cite{Shim2013} and \cite{Miyaji2018}, thereby we cannot restrict the neighbours of each source to a certain velocity interval. The rest of the sources without spec-$z$ had their photo-$z$ calculated. Instead, for each source with photo-$z$ or spec-$z$ (we use spec-$z$ whenever it is available), we select neighbouring galaxies within a certain redshift bin, $\sigma_{\rm bin}$, before calculating the local galaxy density. $\sigma_{\rm bin}$ is calculated using: 
\begin{eqnarray}
    \sigma_{\rm bin} = {\sigma_{\Delta{z/(1+z)}}}(1 + z)
\label{eq:redshiftbin}
\end{eqnarray}
where {$\sigma_{\Delta{z/(1+z)}}$} is the photo-$z$ dispersion of our sources for density calculation. It is calculated using the normalised median absolute deviation (NMAD) \citep{Hoaglin1983}:
\begin{eqnarray}
    \sigma_{\Delta{z/(1+z)}} = 1.48 \times \ {\rm median} \left|\frac{z_{\rm p} - z_{\rm s}}{1+z_{\rm s}}\right|
\label{eq:NMAD}
\end{eqnarray}
where $z_p$ and $z_s$ are the photo-$z$ and spec-$z$, respectively. Rearranging this equation gives the photo-$z$ uncertainty |$z_{\rm p} - z_{\rm s}$| equal to $\sigma_{\rm bin}$. In addition, having the $(1+z)$ factor causes our redshift bins to be redshift dependent (i.e., increasing with redshift). The median value of $\sigma_{\rm bin}$ is 0.114, while the range of $\sigma_{\rm bin}$ is 0.064 to 0.150.

One common issue in using local density as an environmental parameter is edge correction. When a target galaxy is close to the survey edge, it is possible that there are other galaxies close to the target galaxy that is not covered by the survey. Therefore, there is a chance that the measured density is smaller than what it is supposed to be ($\theta_{n}$ is apparently "larger" because the other galaxies that may be closer to it are outside the survey area). Previous studies \citep[e.g.,][]{Miller2003} did "edge correction" by removing "edge galaxies" whose distances from the edge are smaller than their nth-nearest neighbour distance. However, this edge effect may be a problem for surveys with smaller survey areas (e.g., GOODS-North Field with an area of 10$^\prime$ $\times$ 16$^\prime$ or approximately 0.044 deg$^2$ \citep{Giavalisco2004}) and may remove a significant number of valuable sources for analysis \citep{Cooper2005}. \cite{Cooper2005} suggested to recover the edge galaxies by scaling the measured counts by the amount of aperture area covered by the survey area (this was suggested under measuring galaxy environment by counting the galaxies within a fixed aperture size). Despite the fact that our survey area (5.4 deg$^2$) is large enough, we would like to produce a reliable method for edge correction without removing any sources. 

\begin{figure}
	\includegraphics[width=\columnwidth]{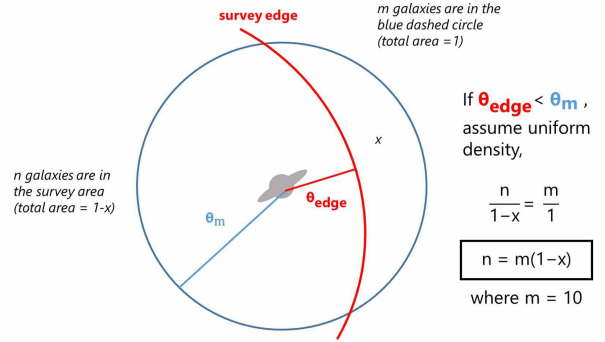}
    \caption{A schematic diagram showing how our edge correction works. The true $m$th nearest neighbour distance ($\theta_{\rm m}$) is the $n$th nearest neighbour distance ($\theta_{\rm n}$). $\theta_{\rm edge}$, on the other hand, is the distance of the source from the survey edge. $x$ is the approximate fraction of the circular area (with $\theta_{\rm m}$ as the radius) that is not covered by the survey. See Sec.~\ref{sec:environment} for a more thorough explanation.}
    \label{fig:edgecorrection}
\end{figure}

Keeping all these in mind, an edge correction method for calculating local galaxy density was devised and implemented to correct the density of edge galaxies. A diagram explaining our method is shown in Fig.~\ref{fig:edgecorrection}. First, "edge galaxies" were selected by identifying galaxies whose distance from the edge ($\theta_{\rm edge}$) is smaller than their $m$th nearest neighbour distance ($\theta_{\rm m}$). Afterwards, we considered a circular area with radius equal to the distance to $\theta_{\rm m}$, and so this circular area must contain $m$ numbers of galaxies. We then measured the approximate area that is not covered by the survey (that is, the part of this circle that is outside the survey edge). Let $x$ be the approximate fraction of the area that is not covered by survey, so that $(1-x)$ of the circular area contains $n$ number of galaxies. Assuming that the number of galaxies is proportional to the area fraction (i.e., uniform density), we obtained the following relationship:

\begin{eqnarray}
    n = m(1-x)
    \label{eq:edgecorrection}
\end{eqnarray}

Eqn.~\ref{eq:edgecorrection} tells us that the true $m$th nearest neighbour distance is the $n$th nearest neighbour distance. In this method, we used the galaxies included in the survey to estimate the edge galaxies' "true" local density. Our work dealt with density related to the 10th nearest neighbour distance, so we set $m=10$. For example, if 40\% of this circular area is not covered by the survey, $n = 6$. Therefore, we considered the 6th nearest neighbour as the galaxy's "true" 10th nearest neighbour. The robustness of this edge correction method is more thoroughly discussed in Appendix \ref{sec:appendixB}, where we tried different values of $n$ aside from $n = 10$. Our robustness tests show that among the possible values of $n$ used, $n=10$ performs well in terms of estimating the correct local density of edge galaxies. Larger values of $n$ may make cluster finding algorithms using our density calculations to be less sensitive to overdensities, while smaller values of $n$ cause our edge correction to be erroneous.

Among the 366791 HSC sources in the NEPW field with density values, 5601 (2$\%$) of them are edge galaxies (with corrected density values). On the other hand, among the 1210 MIR-FIR sources with reliable SED fits and density values at z $\leq$ 1.2, only 10 (1$\%$) of them are edge galaxies. Despite the very low number of edge galaxies in our sample, we believe that our edge correction is still valuable for our research.

\begin{figure}
	\includegraphics[width=\columnwidth]{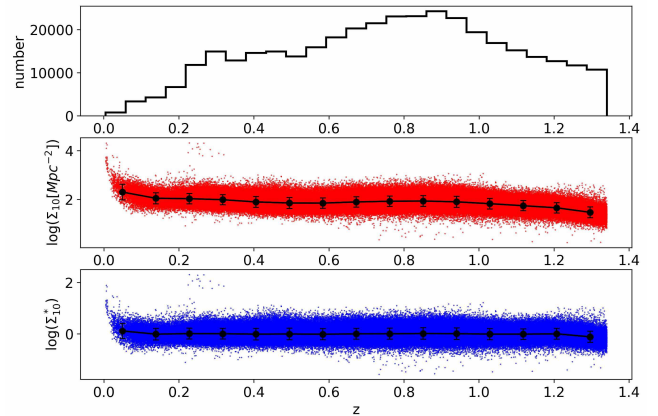}
    \caption{Histogram of 366791 sources with density values (top panel), logarithm of density, $\Sigma_{10}$, in ${\rm Mpc}^{-2}$ (middle panel, red scatterplot) and logarithm of normalised density, $\Sigma^*_{10}$, vs. redshift, $z$. For the scatterplots, the black line with 1$\sigma$ error bars connected in a line shows the median line.}
    \label{fig:normalisation}
\end{figure}

The density values of each source were also normalised by the median density of the redshift bin where it belongs to, making our density estimator unitless. Normalisation was needed to remove the redshift dependence of density, which is shown in Fig~\ref{fig:normalisation}. Without normalisation, the density would decrease with redshift (Pearson rank correlation coefficient is $r$ = -0.37), which we attribute to faint galaxies being less likely to be detected at higher redshifts. However, normalisation alleviates this effect, producing a Pearson rank correlation coefficient of $r$ = -0.05 (both Pearson rank r-coefficient values showed a $p$-value close to zero).

\section{Results}
\label{sec:results}

This section is dedicated to understanding the AGN activity-environment relation in galaxies selected in the \textit{AKARI} NEPW field. We first discuss the distribution of our sample in terms of luminosity and redshift in Sec.~\ref{sample_distribution}. The effect of normalised local density, $\Sigma_{10}^*$, in the AGN activity of our sources is investigated in Sec~\ref{res-env}.

\subsection{Binning of MIR-FIR Sample Distribution}
\label{sample_distribution}

\begin{figure}
    \centering
	\includegraphics[width=\columnwidth]{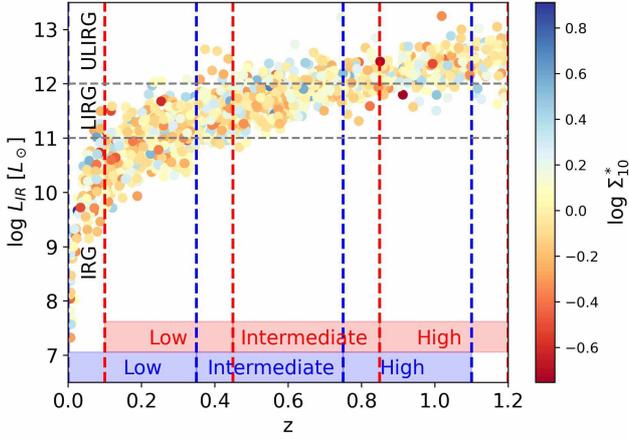}
    \caption{Scatterplot of our sources' logarithm of total IR luminosity vs. redshift. The redshift-luminosity bins used in this work are shown by the dashed lines. Blue (red) dashed vertical lines show the original (shifted) redshift bins. The gray horizontal lines show the luminosity bins. The color of the data points refer to the logarithm of their normalised densities (log $\Sigma_{10}^*$).}
    \label{fig:binning}
\end{figure}

In our analyses, we divided our 1210 all-$z$ sample into different luminosity bins pertaining to different classes of infrared galaxies: (normal) infrared galaxies, hereafter IRGs [$\log ({\rm L}_{\rm TIR}/{\rm L}_\odot) \leq 11$], luminous infrared galaxies, hereafter LIRGs [11 < $\log ({\rm L}_{\rm TIR}/{\rm L}_\odot) \leq 12$], and ultraluminous infrared galaxies, hereafter ULIRGs [12 < $\log ({\rm L}_{\rm TIR}/{\rm L}_\odot)$]. The total IR luminosity of each source were calculated using Eqn.~\ref{eq:agncontributionfrac} by dividing the AGN luminosity by the AGN contribution fraction. Then, for each luminosity bin, we further divided the samples into 3 redshift bins: 0.00 < $z \leq$ 0.35, 0.35 < $z \leq$ 0.70, and 0.70 < $z \leq$ 1.10. However, some of the redshift-luminosity bins had very small number of sources to produce significant statistical results (sample size $\leq$ 10), and so they were excluded from our analyses. To include some of the removed sources in our analyses and investigate how our results change with varying redshift bins, we decided to shift the redshift bins by adding 0.1. The value 0.1 was decided from the photo-$z$ dispersion (0.064), indicating that the photo-$z$ may be uncertain by about 0.1. The resulting shifted redshift bins were: 0.10 < $z \leq$ 0.45, 0.45 < $z \leq$ 0.80, and 0.80 < $z \leq$ 1.20. The reason for binning the sources is to see the dependence of the AGN activity-environment relation on redshift and luminosity. Fig. \ref{fig:binning} shows a visualisation of our redshift-luminosity binning.

\subsection{AGN Activity vs. Normalised Local Density}
\label{res-env}

\begin{figure}
    \centering
	\includegraphics[width=\columnwidth]{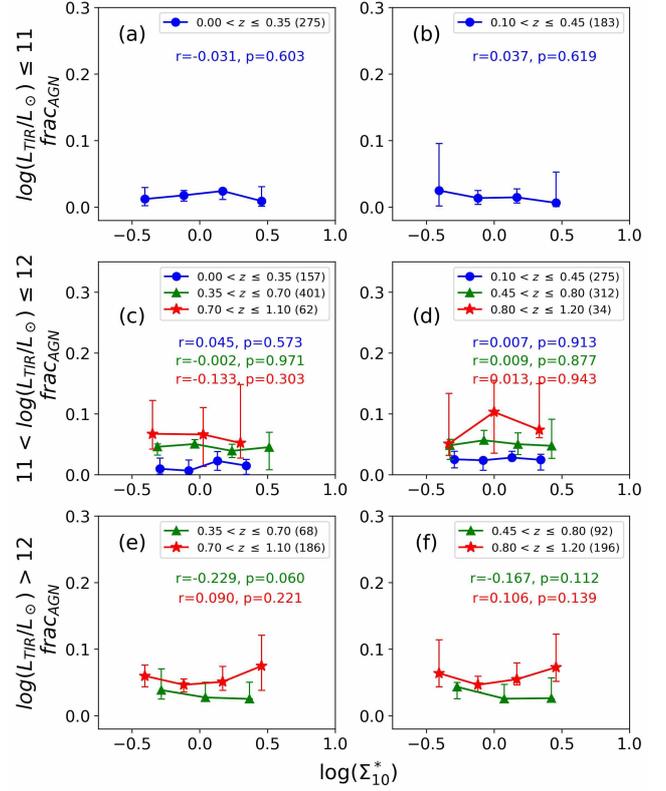}
    \caption{Median lines of AGN contribution fraction (${\rm frac}_{\rm AGN}$) vs. logarithm of normalised local density, log $\Sigma_{10}^*$, for the original redshift bins (left column; panels a, c, and e) and shifted redshift bins (right column; panels b, d, and f). Varying colors and markers indicate redshift bins as shown by the respective legends. Each row represents luminosity groups: IRG (panels a and b), LIRG (panels c and d), and ULIRG (panels e and f). The number of sources per redshift-luminosity bin are enclosed in parenthesis. For the error bars, bootstrapped (10000 times) 95\% confidence intervals (for ${\rm frac}_{\rm AGN}$) were used. The Pearson correlation coefficients ($r$) and p-values ($p$) are also shown in each panel, following the same color coding. The density bins are of equal width and contain at least 15 sources (except for redshift-luminosity bins with $<$ 100 sources, in which case we only make sure that the density bins are of equal width).}
    \label{fig:agn-nld}
\end{figure}

First, we investigated the relationship between AGN contribution fraction ${\rm frac}_{\rm AGN}$ and logarithm of the normalised local density (log $\Sigma_{10}^*$) in Fig.~\ref{fig:agn-nld}. For IRGs and LIRGs, the median line of ${\rm frac}_{\rm AGN}$ as a function of log $\Sigma_{10}^*$ showed very mild to almost no trend for the first two redshift bins, but the highest redshift bin (0.70 < $z \leq$ 1.10 and 0.80 < $z \leq$ 1.20) for LIRGs showed large errors aside from the lack of trend. As a countercheck, we also investigated the Pearson rank correlation coefficients ($r$) and $p$-values of ${\rm frac}_{\rm AGN}$ vs. log $\Sigma_{10}^*$. The Pearson rank correlation coefficient assesses the linear correlation between two datasets (in this case, ${\rm frac}_{\rm AGN}$ and log $\Sigma_{10}^*$) without taking into consideration the uncertainties from the median lines). Our countercheck shows that the $p$-values for IRGs and LIRGs are very large across all redshift bins (not to mention that their Pearson rank correlation coefficients are very close to zero) suggesting that their correlation is not statistically significant. 

However, in Fig.~\ref{fig:agn-nld} we see for ULIRGs that at the intermediate redshift bins (0.35 < $z \leq$ 0.70 and 0.45 < $z \leq$ 0.80), ${\rm frac}_{\rm AGN}$ mildly decreases with log $\Sigma_{10}^*$, but this reverses at the highest redshift bins (0.70 < $z \leq$ 1.10 and 0.80 < $z \leq$ 1.20). The Pearson correlation coefficients for ULIRGs' ${\rm frac}_{\rm AGN}$ vs. log $\Sigma_{10}^*$ imply that these trends are small or weak (|$r$| $<$ 0.3). We also see that their $p$-values are relatively lower compared to those from our IRG and LIRG samples. However, the p-values are still large to confirm the significance of these trends. Choosing our significance level to be $p$ $<$ 0.05 shows that the correlations between ${\rm frac}_{\rm AGN}$ and log $\Sigma_{10}^*$ among different redshift and luminosity bins are not statistically significant.

%with the highest $p$-value recorded for the highest shifted redshift bin (0.80 < $z \leq$ 1.20), p = 0.221, suggesting that the resulting correlations are more likely to be significant.

%\begin{table*}
%\centering
%\caption{Pearson rank correlation coefficient ($r$) and $p$-values of our sample's AGN contribution fraction red(${\rm frac}_{\rm AGN}$) as a function of log $\Sigma_{10}^*$ divided into different redshift bins per luminosity bin. Italicized font indicate values for the shifted redshift bins.}
%\label{tab:rpvalue}
%\begin{tabular}{|c|cc|cc|cc|}
%       \hline
%       & \multicolumn{2}{c|}{0.00 < z $\leq$ 0.35 \textit{(0.10 < z $\leq$ 0.45)}} & \multicolumn{2}{c|}{0.35 < z $\leq$ 0.70 \textit{(0.45 < z $\leq$ 0.80)}} & \multicolumn{2}{c|}{0.70 < z $\leq$ 1.10 \textit{(0.80 < z $\leq$ 1.20)}}  \\
%       \cline{2-7}
%       & $r$ & $p$-value & $r$ & $p$-value & $r$ & $p$-value  \\
%       \hline
%    IRG & -0.031 \textit{(0.307)} & 0.603 \textit{(0.619)} &   &   &   &   \\
%    \hline
%    LIRG & 0.045 \textit{(0.007)} & 0.573 \textit{(0.913)} & -0.002 \textit{(0.009)} & 0.971 \textit{(0.877)} & -0.133 \textit{(0.013)} & 0.303 \textit{(0.943)} \\
%    \hline
%    ULIRG &  &  & -0.229 \textit{(-0.167)} & 0.060 \textit{(0.112)} & 0.090 \textit{(0.106)} & 0.221 \textit{(0.139)} \\
%    \hline
%\end{tabular}
%\end{table*}

\begin{figure}
    \centering
	\includegraphics[width=\columnwidth]{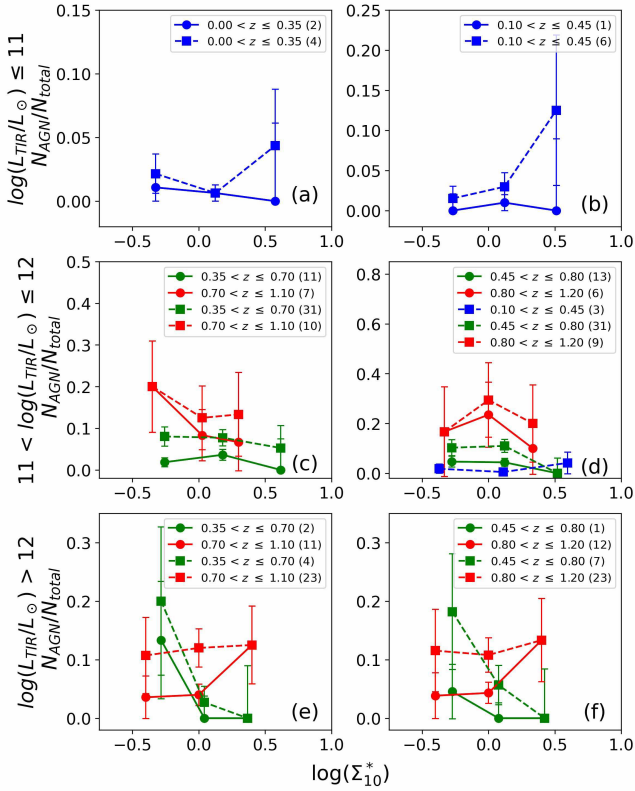}
    \caption{AGN number fraction (${\rm N}_{\rm AGN}/{\rm N}_{\rm tot}$) vs. logarithm of normalised local density, log $\Sigma_{10}^*$, for the original redshift bins (left column; panels a, c, and e) and shifted redshift bins (right column; panels b, d, and f). Solid lines with circle markers refer to sources with ${\rm frac}_{\rm AGN}$ $\geq$ 0.2 selected as AGNs. On the other hand, dashed lines with square markers refer to sources with ${\rm frac}_{\rm AGN}$ $\geq$ 0.15 selected as AGNs. Varying colors indicate redshift bins as shown by the respective legends. Each row represents luminosity groups: IRG (panels a and b), LIRG (panels c and d), and ULIRG (panels e and f). The number of sources per redshift bin for each luminosity group are enclosed in parenthesis. For the error bars, Poisson error bars were used. The density bins are of equal width and contain at least 15 sources (except for redshift-luminosity bins with $<$ 100 sources, in which case we only make sure that the density bins are of equal width). For ${\rm N}_{\rm AGN}/{\rm N}_{\rm tot} =$ 0, Poisson single-sided upper limits corresponding to 1$\sigma$ \citep{Gehrels1986} are shown.}
    \label{fig:agn_number-nld}
\end{figure}

We also investigated the AGN number fraction (${\rm N}_{\rm AGN}/{\rm N}_{\rm tot}$) as a function of log $\Sigma_{10}^*$ in Fig~\ref{fig:agn_number-nld}. We first considered all sources with ${\rm frac}_{\rm AGN}$ $\geq$ 0.2 as AGNs (solid lines). One limitation of this criterion is that it selects very few sources for our analysis, contributing to the large error bars (Poisson error) across all redshift bins for all luminosity groups. However, the increasing (decreasing) trend of AGN activity (in this case, AGN number fraction) is still apparent for the highest (intermediate) redshift bin for ULIRGs. We also tried to lower the ${\rm frac}_{\rm AGN}$ threshold of selecting AGNs from 0.2 to 0.15 (dashed lines). Fig.~\ref{fig:agn_number-nld} shows the result of adjusting this threshold. As far as the ULIRGs are concerned, the aforementioned trends are still apparent, but a larger sample of AGN further emphasised these trends.

\section{Discussion}
\label{sec:discussion}

\subsection{Effect of Stellar Mass}
\label{sec:stellar_mass}

\begin{figure}
    \centering
	\includegraphics[width=\columnwidth]{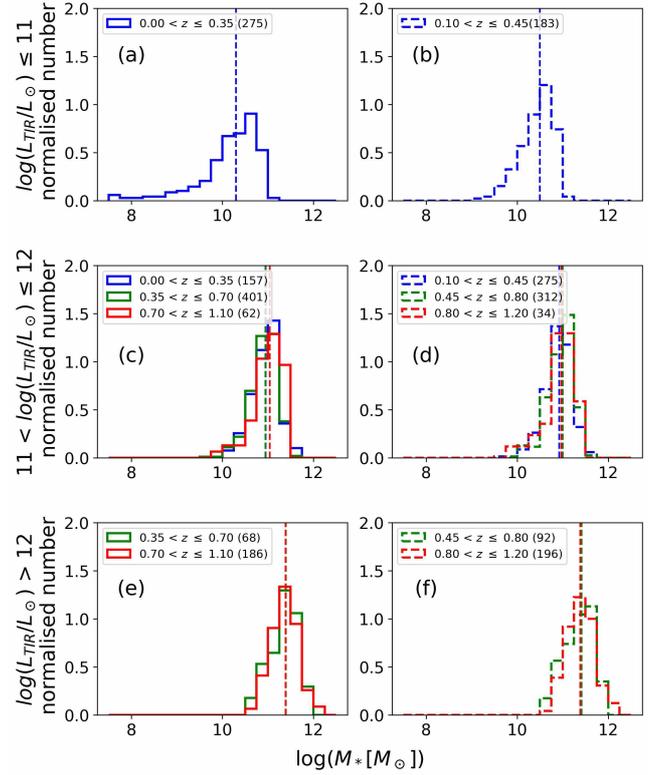}
    \caption{Histogram of log $M_* [M_\odot]$ for the original redshift bins (left column; panels a, c, and e) and shifted redshift bins (right column; panels b, d, and f). Each row represents luminosity groups: IRG (panels a and b), LIRG (panels c and d), and ULIRG (panels e and f). The number of sources per redshift bin for each luminosity group are enclosed in parenthesis. The vertical dashed lines correspond to the median log $M_* [M_\odot]$ of the distributions.}
    \label{fig:stellar_mass}
\end{figure}

In this section, we discuss the effect of stellar mass on our results, especially in the suggested reversal trend observed in the SFR-density relation for our ULIRG sample. Since our results suggest a reversal in the relationship between AGN activity and environment in ULIRGs, we need to check the stellar mass distributions of our sources at different redshift-luminosity bins. Fig~\ref{fig:stellar_mass} shows the histogram of the sources' logarithm of stellar mass, log $M_* [M_\odot]$. As expected, increasing luminosity implies increasing stellar mass distributions. For each luminosity group, we also performed a Kolmogorov-Smirnov (KS) test as a countercheck. For LIRGs, we found relatively significant differences for pairs of stellar mass distributions. For instance, we have $p$ $\approx$ 0.018 for the stellar mass distributions of LIRGs in the highest original redshift bin (0.70 $<$ $z$ $\leq$ 1.10) and the intermediate original redshift bin (0.35 $<$ $z$ $\leq$ 0.70). We also have $p$ $\approx$ 0.063 for the stellar mass distributions of LIRGs in the highest original redshift bin and lowest original redshift bin (0.00 $<$ $z$ $\leq$ 0.35). For the shifted redshift bins, the LIRGs at the lowest (0.10 $<$ $z$ $\leq$ 0.45) and intermediate (0.45 $<$ $z$ $\leq$ 0.80) shifted redshift bins showed significant difference ($p$ $\approx$ 0.003). The rest have very high $p$-values ($p$ $\geq$ 0.23). However, our ULIRGs' did not show any significant differences in their stellar mass distributions even at the shifted redshift bins ($p$ $\geq$ 0.29). The mixed results of significant differences in the stellar mass distributions of our LIRGs may have caused the absence of apparent correlation between AGN activity and environment. However, for ULIRGs, the effect of stellar mass is not so strong or has no clear evidence of affecting our results.

\begin{figure}
    \centering
	\includegraphics[width=\columnwidth]{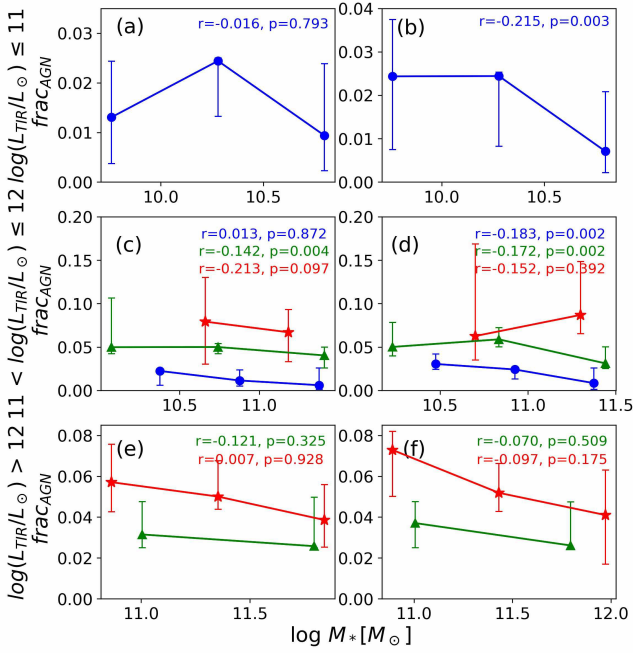}
    \caption{Similar to Fig. \ref{fig:agn-nld}, but for ${\rm frac}_{\rm AGN}$ vs. log $M_* [M_\odot]$. The same legends apply.}
    \label{fig:agn-mass}
\end{figure}

\begin{figure}
    \centering
	\includegraphics[width=\columnwidth]{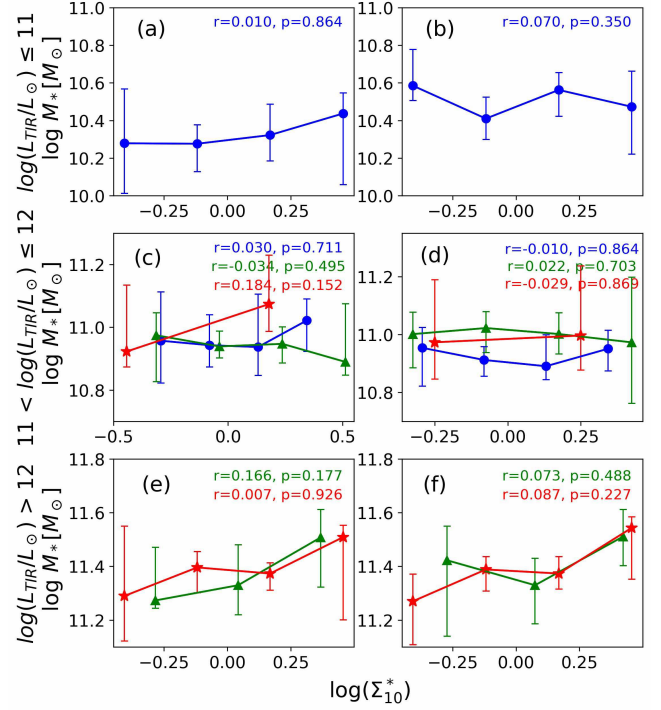}
    \caption{Similar to Fig. \ref{fig:agn-nld}, but for log $M_* [M_\odot]$ vs. log $\Sigma_{10}^*$. The same legends apply.}
    \label{fig:mass-nld}
\end{figure}

We also investigated the relationship between stellar mass and density/AGN activity. Previous studies \citep{Elbaz2007, Hwang2019} showed a reversal in SFR-density relation at higher redshifts (increasing SFR with increasing density in contrast with lower redshifts). They explored the effect of stellar mass in their observed reversal trend. In \cite{Elbaz2007}'s work, if SFR increases with stellar mass at higher redshifts and stellar mass increases with density at higher redshifts, stellar mass may serve as a physical reason why there is enhanced SF activity in denser environments at higher redshifts. In our work, instead of looking at SFR, we explore the correlation between ${\rm frac}_{\rm AGN}$ and log $M_*$, and stellar mass and log $\Sigma_{10}^*$. These are shown in Figs. \ref{fig:agn-mass} and \ref{fig:mass-nld}, respectively. Although our small sample size limits us to reach high significance levels ($p$ $<$ 0.05) in most of our redshift-luminosity bins, it is quite clear that just by looking at our ULIRG sample, their ${\rm frac}_{\rm AGN}$ decreases with stellar mass (which is also suggested in the other luminosity bins), and stellar mass increases with log $\Sigma_{10}^*$ (which is less marginal in other luminosity bins) across intermediate and high redshift bins, regardless of shifting them or not. This implies that we should expect that ${\rm frac}_{\rm AGN}$ decreases with log $\Sigma_{10}^*$ across all redshift bins as far as our ULIRG sample is concerned. However, this is not what our results show (panels (e) and (f) in Fig. \ref{fig:agn-nld}), as we see that ${\rm frac}_{\rm AGN}$ increases with $\Sigma_{10}^*$ at our highest redshift bin. Our results suggest that there is a possibility of environmental dependence of AGN activities in ULIRGs.

\begin{figure}
    \centering
	\includegraphics[width=\columnwidth]{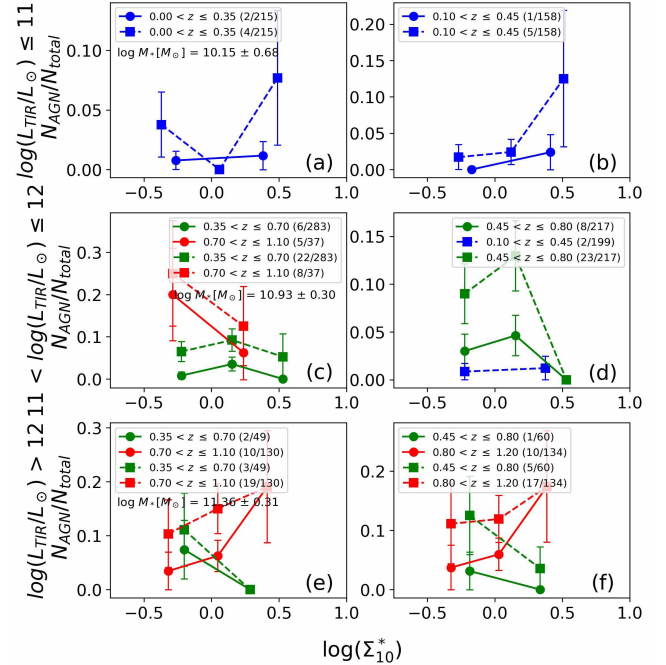}
    \caption{Similar to Fig. \ref{fig:agn_number-nld}, but for each luminosity bin (each row), the stellar mass of the sample are narrowed down (the stellar mass ranges for each luminosity bin are shown in the first column). In the legends of each panel, the number of AGNs over the new total number of sources per redshift-luminosity bins are shown. The same legends apply.}
    \label{fig:agn_number-nld_narrow}
\end{figure}

\cite{Hwang2019}, on the other hand, looked at the possibility of the observed SFR-density reversal being an effect of galaxies in different density bins having different stellar masses. To verify this, they narrowed the mass range of the sample galaxies and tried to observe the same reversal trend. We implored the same method in Fig. \ref{fig:agn_number-nld_narrow}, wherein we narrowed down the mass range of our luminosity bins by selecting galaxies within the average log $M_* [M_\odot]$ $\pm$ the standard deviation of log $M_* [M_\odot]$ for each luminosity bin. It is clear from Fig. \ref{fig:agn_number-nld_narrow} that even if restrict our sample into narrow stellar mass bins, the reversal in AGN number fraction-environment relation for ULIRG remains unchanged. We therefore conclude that stellar mass does not have a strong effect in the suggested reversal trend in ULIRG AGN activity-environment.

\subsection{Selection bias due to FIR detection}
\label{sec:selection_bias}

The FIR detection, as also explained by \cite{Wang2020}, also produces selection bias for our study. As explained earlier, this criterion is important to make sure that our SED fitting is robust via the energy balance principle. The resulting selection bias happens because this criterion will make the sources flux-limited in FIR at higher redshift. This also means that we can only select very luminous FIR galaxies at high redshift, and high-$z$ objects dominated by AGNs but lack far-IR detection might be dropped from our sample due to this criteria. We therefore expect that majority of our galaxies have high star-formation activity. However, because of the multiwavelength data that we have and the SED fitting method that we utilised, we are able to detect their obscured AGN activity. As a comparison, \cite{Yang2020} in their sample of X-ray selected \textit{AKARI} NEPW sources showed an AGN contribution fraction ranging from 0.2 - 0.7.

\subsection{Difference between our work and previous works}
\label{sec:difference_works}

There have been very few studies which focused on how galaxy environment affects AGN activity based on the sample selection in the infrared wavelengths (especially based on the MIR detection), as the AGN activity-environment research field is dominated by radio, optical, and X-ray detected sources. This is because it has been difficult to obtain a significantly statistical sample in IR (particularly MIR) until this work.

Based on \cite{Malek2017}'s sample of LIRGS and ULIRGs, \cite{Bankowicz2018} studied 22 LIRGs and 17 ULIRGs in 0.06 $\leq$ $z$ $\leq$ 1.23 from the \textit{AKARI} Deep Field-South (ADF-S) survey (observed in 4 \textit{AKARI} FIR filters) and in other public databases (e.g., IRSA, SIMBAD, NED). They used \texttt{CIGALE} SED Fitting to constrain the galaxy properties of their sample, and their density values were calculated using optical images from Digitised Sky Survey (DSS). Their results, show that LIRGs do not have a correlation between AGN contribution fraction and local density (similar to ours), but ULIRGs' AGN contribution fraction increases with environmental density (contrary to ours). While they only focused on FIR-detected sources, we focused on sources that were both detected in MIR and FIR. These wavelength ranges probe different processes. MIR probes the dust heated by AGN, while FIR probes the cold interstellar medium (ISM) dust. Moreover, our work used multiwavelength data with at most 36 bands for constraining galaxy properties, which is much more than that of \cite{Bankowicz2018}'s (at most 24 bands).  The number of sources in our study is 2 orders of magnitude larger than that of \cite{Bankowicz2018}. Therefore, there is an improvement in the sample size and quality of our data. Lastly, they were not able to divide the LIRGs and ULIRGs into different redshift bins. In our work, we emphasize the importance of doing this to disentangle the role of environment, redshift, and galaxy properties (in this case, total IR luminosity) on AGN activity. We were able to show that at different redshift bins, ULIRGs exhibit both increasing and decreasing trends between AGN contribution fraction and density. 

\cite{Hickox2009} focused on AGN populations detected in various wavelengths: radio AGNs (with 1.4 GHz detection by the Westerbork Synthesis Radio Telescope), X-ray AGNs (detected in the \textit{Chandra} XBo{\"o}tes Survey), and MIR (detected in the $Spitzer$ IRAC Shallow Survey). %Their results with dark matter halo masses derived from spatial two-point correlation analysis show that 
Their radio AGN sample are the most strongly clustered, followed by X-ray AGNs, and then MIR AGNs. Their results also imply that weak clustering occurs in slightly bluer and less luminous galaxies. \cite{Karouzos2014}, on the other hand, studied the possibility of mergers as triggers of dust-obscured MIR AGNs selected in VISTA Deep Extragalactic Observations (VIDEO) survey of XMM-Large Scale Structure Field. By calculating the number of galaxies within a set aperture size, finite redshift slice and local galaxy density, they showed that MIR selected AGNs do not prefer dense close environments (approximately 2 arcmin, where companion galaxies may be found) over cluster environments. Still, their samples exhibit significant uncertainties, which may be due to the fact their sample is contaminated by many SFGs as their sample selection also relies on MIR detection (which is also the case in our work). Finally, \cite{Martini2013} suggested a reversal in the relationship between AGN number fraction and cluster-field environment. By utilising galaxy clusters selected from $Spitzer$/IRAC Shallow Cluster Survey, they found out X-ray AGN number fraction is lower in clusters than in field environments at higher redshift (1 < $z$ < 1.5). However, X-ray and MIR AGN number fractions are consistent for both cluster and field environments at low redshift ($z$ < 1) \citep{Martini2009}, indicating an evidence of environment-dependent evolution of AGN only for X-ray AGNs. 

\cite{Magliocchetti2018b} mainly focused on radio-detected AGNs, and those with counterparts from $Herschel$ (FIR detections). They found that radio AGNs with MIR detections prefer field environments at redshift $z$ $\leq$ 1.2 and are found in low-mass galaxies. \cite{Klesman2014} also investigated the radial distribution of multiple AGN populations in clusters selected in the Hubble Space Telescope (HST) Advanced Camera for Surveys (ACS) in optical, X-ray, and MIR wavelengths. They found that MIR AGNs are less centrally concentrated (less galaxy density at the cluster center) compared to X-ray AGNs, which are shown to be much more centrally concentrated than normal galaxies within the 20\% of the cluster virial radius. An earlier work by \cite{Klesman2012} using the same cluster sample also showed that there is no significant difference between AGN number fraction among these clusters and AGNs detected in the Great Observatories Origins Deep Survey (GOODS) fields. 

However, there are also studies that showed that MIR AGN activity prefer dense environments. For instance, \cite{Galametz2009} investigated the X-ray, MIR, and radio AGNs in clusters within the redshift range 0 < $z$ < 1.5 within the Bootes field of NOAO Deep Wide-Field Survey. In their work, MIR AGNs do not show any significant preference in $z$ < 0.5 clusters, but for clusters in 0.5 < $z$ $\leq$ 1.0, a weak overdensity of MIR AGNs is shown at the cluster centers (< 0.3 Mpc). In addition, \cite{Krick2009} studied 3 MIR clusters detected in the IRAC Dark Field at $z$ = 1, and the AGN number fraction (and summed SFR) of cluster galaxies increases with redshift faster than that of fields, suggesting an indirect environmental effect on AGN activity of their sample.

Based on these works, we can clearly see that there is a variety of conclusions drawn from different MIR AGN activity - environment studies, which is partly due to different sample selection criteria and choice of environmental parameters. However, some of these works' conclusions are similar to ours. For instance, we found no significant correlation between AGN activity and environment for most of our samples (IRGs and LIRGs), which is similar to \cite{Karouzos2014} and \cite{Bankowicz2018}. This can be attributed to the selection bias caused by our sample selection criteria in the MIR and FIR wavelength regions.

As for other wavelengths, \cite{Georgakakis2008} showed that X-ray AGNs at 0.7 < $z$ < 1.4 prefer group environments at 99\% confidence level. They attributed this to the fact that AGNs are preferentially hosted by red luminous galaxies which usually reside in dense environments. As they factored in the host galaxy properties, they showed that this significance drops to 91\%, indicating that X-ray AGNs live in diverse environments. The importance of factoring the host galaxy properties were also emphasized by \cite{Silverman2009}. Their work showed that there is a lack of environmental dependence on X-ray AGN activity over galaxies with lower stellar masses (10.4 < log $M_*$ < 11) selected in the zCOSMOS spectroscopic redshift survey. However, for galaxies with stellar masses log $M_*$ > 11, strong X-ray AGNs were found to prefer lower-density environments and live in more massive and bluer host galaxies. They concluded that environmental effects on AGN activity is dependent on the host galaxy properties such as stellar and/or gas mass content, and so overall AGNs prefer environments similar to that of massive galaxies with ample amounts of star formation.

There are also works that showed reversal of environmental effects on SFR and fraction of early-type galaxies instead of AGN activity. For instance, \cite{Hwang2019} used cosmological hydrodynamical simulation of galaxies within 0.0 $\leq$ $z$ $\leq$ 2.0 to show that at $z$ $\leq$ 1, SFR increases with local density instead of decreasing. According to their work, massive star-forming galaxies, which have large amounts of cold gas for SF activity, are highly clustered at high redshift. Because of environment-induced activities, these galaxies consume cold gas much faster at high-density regions compared to low-density regions. As a result, massive star-forming galaxies are found at high-density regions and high redshift, but at low redshifts, these galaxies end up becoming quiescent in high-density regions. A reversal in galaxy morphology transformation with environment at different redshifts was also shown by \cite{Hwang2009}. Using spectroscopically observed galaxies at 0.0 $\leq$ $z$ $\leq$ 1.0 in the Great Observatories Origins Deep Survey (GOODS), the correlation between galaxy morphology and local density is weaker at higher redshifts than at low redshifts. This implies that frequent galaxy-galaxy interactions at high-density regions at high redshifts cause the morphological transformation of late-type galaxies into early-type galaxies to accelerate together with the quenching of their SF activity.

\subsection{Monte Carlo Simulations of ULIRGs' AGN activity - Local Environment relation}
\label{sec:monte-carlo}

We showed that ULIRG AGN number fraction increases with log $\Sigma_{10}^*$ at the highest redshift bin (0.70 < $z$ $\leq$ 1.10 and 0.80 < $z$ $\leq$ 1.20), but the opposite is observed for ULIRGs at the intermediate redshift bin (0.35 < $z$ $\leq$ 0.70 and 0.45 < $z$ $\leq$ 0.80). We do not see a clear trend on the IRG and LIRG AGN number fraction and log $\Sigma_{10}^*$, and the observed trends remain unchanged despite lowering the ${\rm frac}_{\rm AGN}$ threshold in selecting AGNs from 0.20 to 0.15. Investigating the ${\rm frac}_{\rm AGN}$ vs. log $\Sigma_{10}^*$ at different redshift-luminosity bins also shows similar trends with the AGN number fraction. We recovered large p-values for IRGs and LIRGs ($p \gg 0.5$), while p-values for ULIRGs are relatively low (p $\leq$ 0.221). However, we cannot rule these trends as significant as they do not fall under our significance level ($p$ $<$ 0.05). Visually, our results still suggest a reversal in the ULIRG AGN activity-environment relation at different redshift bins.

\begin{figure}
    \centering
	\includegraphics[width=0.8\columnwidth]{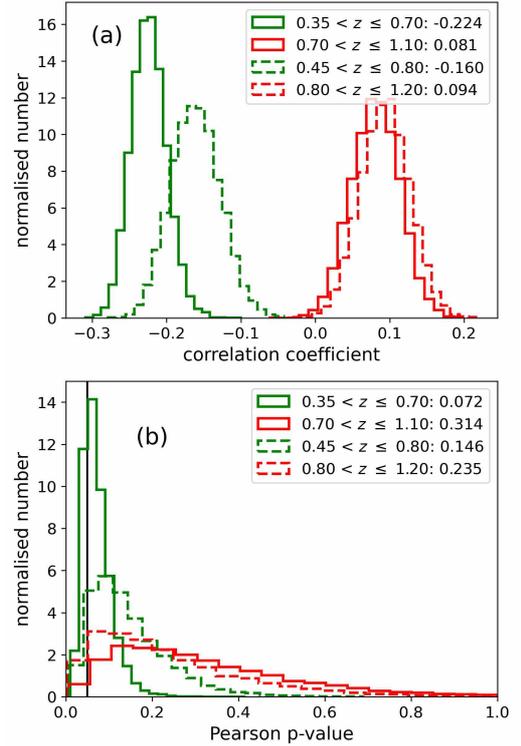}
    \caption{The distribution of the Pearson correlation coefficients (upper panel, a) and p-values (lower panel, b) for ULIRG ${\rm frac}_{\rm AGN}$ vs. log $\Sigma_{10}^*$. The distributions are calculated from 10,000 Monte Carlo realisations. The ${\rm frac}_{\rm AGN}$ errors are used for $\sigma$ of the Gaussian probability distribution functions. Different colors show different redshift bins, solid (dashed) lines show the original (shifted) redshift bins as shown in the legends. A correlation coefficient close to 1 indicates a strong positive correlation, while a correlation coefficient close to -1 indicates a strong negative correlation. For the p-value distribution (lower panel, b), the solid black vertical line represents $p$=0.05, which is our threshold of significance. The average correlation coefficients and p-values for each redshift bin are shown in the legends.}
    \label{fig:monte_carlo_fracAGN}
\end{figure}

We aim to test the reliability of this reversal trend by running a Monte Carlo simulation for the ULIRG AGN activity-environment relation. The errors of individual data points are considered by adding random errors that follow a Gaussian distribution. This distribution is centered on the actual value, and its standard deviation is similar to the error of the data point. 10,000 errors were simulated and for each instance, we calculated the Pearson coefficients and p-values. We investigated the relation in AGN contribution fraction and AGN number fraction separately. For the AGN contribution fraction (${\rm frac}_{\rm AGN}$), we use the error of the AGN contribution fraction estimated by \texttt{CIGALE} as the error of each data point. Fig. \ref{fig:monte_carlo_fracAGN} shows the result of our Monte Carlo simulation for ${\rm frac}_{\rm AGN}$. The correlation coefficients (upper panel of Fig. \ref{fig:monte_carlo_fracAGN}) at intermediate (0.3 $\lesssim$ $z$ $\lesssim$ 0.7) and high (0.7 $\lesssim$ $z$ $\lesssim$ 1.2) redshift situate at the negative and positive side of the plot, respectively, which suggests the reversal in these redshift bins. The average correlation coefficients, however, indicate marginal correlation only. The p-values indicate the statistical significance of the correlation. As shown in the lower panel of Fig. \ref{fig:monte_carlo_fracAGN}, only a small fraction of the iterations show $p$ $<$ 0.05 (corresponds to 2$\sigma$ significance): for ULIRGs, an average of $\sim$17\% ($\sim$5\%) of the iterations at the intermediate (high) redshift bins have $p$ $<$ 0.05. The averages of the $p$-value distributions for ULIRGs' redshift bins are relatively lower compared to other luminosity groups (the coefficient and p-value distributions of other luminosity groups are included in the Supplementary Material). Therefore, we cannot say with enough confidence that the reversal is statistically significant.

\begin{figure}
    \centering
	\includegraphics[width=0.8\columnwidth]{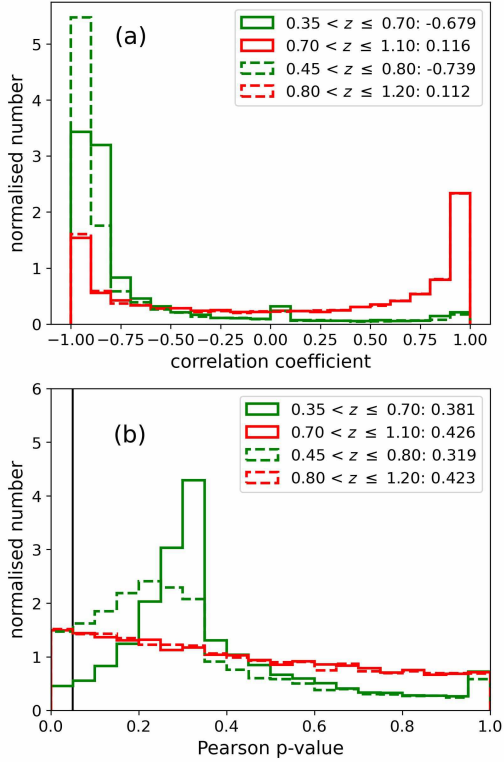}
    \caption{Similar as Fig. \ref{fig:monte_carlo_fracAGN} but with ${\rm N}_{\rm AGN}/{\rm N}_{\rm tot}$ instead of ${\rm frac}_{\rm AGN}$. The error bars in each ${\rm N}_{\rm AGN}/{\rm N}_{\rm tot}$ data point are used for $\sigma$ of the Gaussian probability distribution functions. We only show results with ${\rm frac}_{\rm AGN}$ $\geq$ 0.15 as our AGN selection criteria. The same legends apply.}
    \label{fig:monte_carlo_AGNnumberfraction}
\end{figure}

We also performed Monte Carlo Simulations for the AGN number fraction (${\rm N}_{\rm AGN}/{\rm N}_{\rm tot}$). In this case, the error bars in each data point in Fig. \ref{fig:agn_number-nld} serve as the dispersion of the error distribution. Fig. \ref{fig:monte_carlo_AGNnumberfraction} shows the results of our Monte Carlo Simulations for ${\rm N}_{\rm AGN}/{\rm N}_{\rm tot}$. Due to the sparse number of AGNs in each redshift-luminosity bin as the stricter AGN definition is used (${\rm frac}_{\rm AGN}$ $\geq$ 0.20), we only show the results of our Monte Carlo Simulations using ${\rm frac}_{\rm AGN}$ $\geq$ 0.15, which enables us to select more AGNs yet still shows the same reversal trend in ULIRGs. The correlation coefficient distributions, although on average show similar trends as Fig. \ref{fig:monte_carlo_fracAGN}, are situated in very wide ranges of values. The small number of data points in Fig. \ref{fig:agn_number-nld} limits us to achieve significance in the reversal of the ULIRG AGN number fraction-environment. For ULIRG AGNs with ${\rm frac}_{\rm AGN}$ $\geq$ 0.15, an average of $\sim$5\% ($\sim$8\%) of the iterations at the intermediate (high) redshift bins have $p$ $<$ 0.05 (corresponds to 2$\sigma$ significance). Again, the p-value distributions on average for ULIRGs across all redshift bins are slightly lower compared to the p-value distributions of other luminosity groups. The coefficient and p-value distributions of other luminosity groups for AGN number fraction are included in the Supplementary Material.

%\textcolor{red}{The rest of the Monte Carlo simulation results for other luminosity groups and for ${\rm frac}_{\rm AGN}$ $\geq$ 0.20 as AGN selection criteria are included in the Supplementary Material.}

%local environment may affect the AGN activity of ULIRGs at different redshift bins: at our highest redshift bin (0.70 < $z$ $\leq$ 1.10 and 0.80 < $z$ $\leq$ 1.20), AGN activity (i.e., ${\rm frac}_{\rm AGN}$ and ${\rm N}_{\rm AGN}/{\rm N}_{\rm tot}$) increases with density, but for our intermediate redshift bin (0.35 < $z$ $\leq$ 0.70 and 0.45 $z$ $\leq$ 0.80), AGN activity decreases with density (Figs. \ref{fig:agn-nld} and \ref{fig:agn_number-nld}). \textcolor{red}{We note that the observed trends on ${\rm frac}_{\rm AGN}$ produce large Pearson correlation p-values, and so must be }

\subsection{ULIRGs' AGN activity vs. Redshift in different environment bins}
\label{sec:env_ulirgs}

We further investigated the suggested reversal in AGN activity-environment of our ULIRG sample by plotting ${\rm frac}_{\rm AGN}$ vs. redshift for ULIRGs with high and low density. We define low (high) density ULIRGs if their log $\Sigma_{10}^*$ is less than (greater than) the average log $\Sigma_{10}^*$ (symbolised as $\mu$) of our ULIRG sample. In our analysis, $\mu$ = 0.022. The idea of dividing the ULIRGs based on their density is to investigate the interaction between AGN contribution fraction and redshift of these two groups. \cite{Wang2020} already showed that AGN activity increases with redshift across all luminosity bins/groups. However, based on our initial results, it is possible that at relatively low redshift ranges, AGN activity is high for low density ULIRGs, while the opposite is true for high density ULIRGs.

\begin{figure}
	\includegraphics[width=\columnwidth]{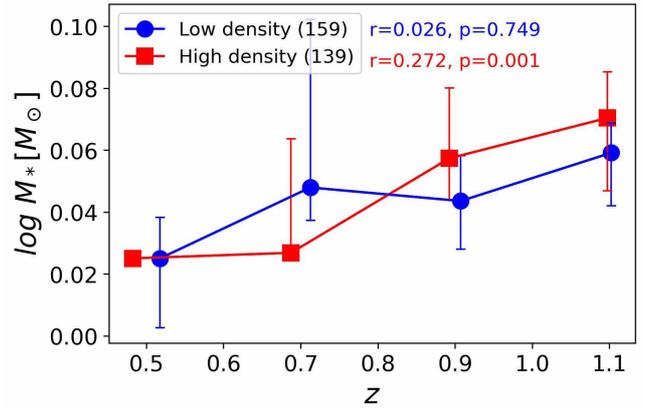}
    \caption{Median line of ${\rm frac}_{\rm AGN}$ vs. $z$ for low density (blue solid line with circle markers) and high density (red dashed line with square markers) ULIRGs. Error bars are bootstrapped 95\% confidence intervals.}
    \label{fig:ULIRGs_fracAGN_density}
\end{figure}

\begin{figure}
	\includegraphics[width=\columnwidth]{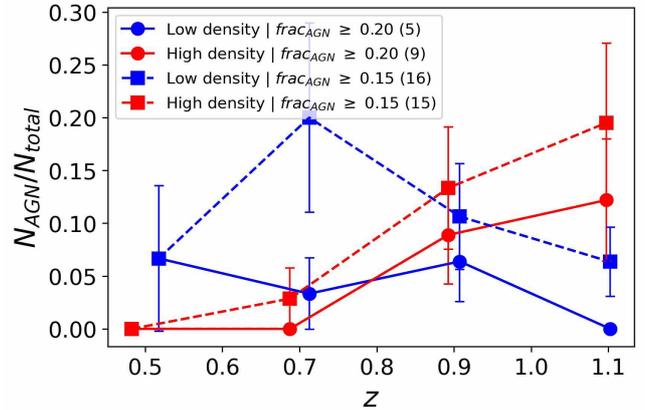}
    \caption{${\rm N}_{\rm AGN}/{\rm N}_{\rm tot}$ vs. $z$ for low density (blue lines) and high density (red lines) ULIRGs. Solid lines with circle markers refer to sources with ${\rm frac}_{\rm AGN}$ $\geq$ 0.2 selected as AGNs. On the other hand, dashed lines with square markers refer to sources with ${\rm frac}_{\rm AGN}$ $\geq$ 0.15 selected as AGNs. Error bars are Poisson error bars.}
    \label{fig:ULIRGs_AGNnumberfrac_density}
\end{figure}

Fig.~\ref{fig:ULIRGs_fracAGN_density} shows the median plot for ${\rm frac}_{\rm AGN}$ vs. redshift ($z$) for the low and high density ULIRGs. At face value, the figure indicates that ${\rm frac}_{\rm AGN}$ increases slowly for low density ULIRGs, but ${\rm frac}_{\rm AGN}$ increases much faster for high density ULIRGs beyond $z$ $\geq$ 0.6.  However, we cannot rule out with ample significance that this is the case due to the large error bars in the plot, which is attributed to the large scatter in our sample. As a countercheck, we report the Pearson rank correlation coefficients and p-values of ${\rm frac}_{\rm AGN}$ vs. $z$: $r$ = 0.026 ($p$ = 0.749) for low density ULIRGs, and $r$ = 0.272 ($p$ = 0.001) for high density ULIRGs. In Fig.~\ref{fig:ULIRGs_AGNnumberfrac_density}, we investigate how ${\rm N}_{\rm AGN}/{\rm N}_{\rm tot}$ changes with redshift for low and high density ULIRGs. Again, despite the relatively large error bars in the figure, it is apparent that ${\rm N}_{\rm AGN}/{\rm N}_{\rm tot}$ increases (decreases) with redshift for high (low) density sources even if we change the AGN selection criteria to a less stricter definition. Figs.~\ref{fig:ULIRGs_fracAGN_density} and \ref{fig:ULIRGs_AGNnumberfrac_density} also suggest that AGN activity-environment trend for ULIRGs reverses at around $z$ $\approx$ 0.8.

\subsection{Possible Explanations for Reversal of AGN activity-environment trend for ULIRGs}
\label{sec:environment_agn}

The main result of our work shows that in general, MIR AGN activity does not depend greatly on galaxy environment. However, our results suggest that AGN activity of ULIRGs may be affected by local environment (defined by normalised local density. We suggest that the effect of local environment on ULIRGs' AGN activity reverses at around redshift $z$ $\approx$ 0.80, although we are limited to the statistics of our sources as these trends show to be statistically insignificant.

There are possible explanations for the suggested reversal in the AGN activity-environment of our ULIRG sample. One possibility is that the effect of environment on MIR AGNs depends on the host galaxy properties, particularly luminosity (or galaxy mass). Previous studies also showed similar trends for more massive galaxies and/or particular classifications of galaxies but for different AGN populations. For instance, \cite{Lopes2017} showed with their sample of cluster galaxies at $z$ $\leq$ 0.1 that optically-selected AGNs prefer field environments more than cluster galaxies especially for more massive galaxies (log $M_*$ [$M_\odot$] > 10.6). \cite{Pimbblet2013} also showed with their sample of SDSS galaxy clusters at $z$ $\approx$ 0.07 that less massive galaxies (log $M_*$ [$M_\odot$] < 10.7) do not show strong environmental variation in their AGN fraction. \cite{Miraghaei2020} studied radio and optical AGNs in SDSS DR7 main galaxy sample at $z$ $\leq$ 0.1 and their environments (whether they are the brightest galaxy in the group, member of a group, or they reside in void or isolated environments). Their results showed that different AGN populations exhibit different environmental effects on their AGN acitivity depending on the host galaxy properties (in their case, they focused on galaxies of different colors, whether they are red, green, or blue galaxies). For instance, optical AGN activity in blue galaxies does not depend on environment, but a higher optical AGN number fraction was found in red galaxies in voids compared to denser environments. Note that these studies focused on optical AGNs, and on lower redshift (local) galaxies. Nevertheless, our speculation about the effect of host galaxy properties in the environmental effect of AGN activity of our MIR sample still holds. This can explain why we can only see significant trends for ULIRGs only.

The suggested reversal in the ULIRGs' AGN activity with local density for different redshift ranges might be indicative of different environmental effects in ULIRGs' AGN activity under different epochs. In their review about ULIRGs, \cite{Lonsdale2006} mentioned that high and low redshift ULIRGs may have formed differently and the environment could affect their growth and properties. This also connects to our previous hypothesis about the possibility that host galaxy properties may affect how galaxy environment affect AGN activity of our MIR sample. Overall, we were able to show that taking into consideration other galaxy properties such as redshift and IR luminosity may also help us speculate more on the true relationship between AGN activity and galaxy environment.

Nevertheless, our results are consistent with the picture that MIR AGN activity does not greatly depend on local galaxy environment. Previous work that focused on cluster-field classification as the main environmental parameter \citep[e.g.,][]{Martini2009, Klesman2012} showed similar results. However, local and global environmental parameters probe galaxy environment differently. Our work is one of the first studies that probed the effect of local galaxy environment on MIR AGN activity, and so it is crucial to expand this study.

\subsection{Limitations and Future Prospects}
\label{sec:limitation_future}

The lack of statistical significance in our results could either be due to the small number of sources in each redshift-luminosity bin, or due to the actual lack of correlation between the two quantities. To verify, future space missions that are able to observe more sources for studying the role of galaxy environment in triggering or suppressing MIR AGN activity are of utmost importance. For instance, the James Webb Space Telescope (\textit{JWST}; \cite{Gardner2006}, will conduct deep observations of the NEPW field in the near future. \textit{AKARI} and \textit{JWST} both employ continuous mid-IR imaging, and so \textit{JWST} will greatly help in increasing our galaxy sample while still benefiting in the continuous mid-IR imaging which is the main advantage of our work. Other future space missions such as \textit{Euclid} \citep{Laureijs2011}, and \textit{SPHEREx} \citep{Dore2016, Dore2018} are also expected to produce great synergy with our \textit{AKARI} data in the NEPW. In connection to having a small number of AGNs in each redshift-luminosity bin, small number statistics \citep{Cash1979} should also be an area for future research. Previous studies have utilised this method in analysing X-ray AGNs (or AGN number fractions) when average AGN spectral parameters can be produced with a (population synthesis) model to calculate the expected number of X-ray AGNs detected, and/or when spectral counts of X-ray sources are lower than the required threshold for statistical analysis \citep[e.g.,][]{Akylas2019, Perola2004}.

Because of the low number of spectroscopically identified sources in the \textit{AKARI} NEPW field, we are limited to photo-$z$ measurements. A larger sample of sources with spec-$z$ can improve the photo-$z$ estimation and density calculation. Therefore, large spectroscopic programs targeting the \textit{AKARI} NEPW field will be helpful for our goal. The Subaru Prime Focus Spectrograph (PFS; \cite{Tamura2016}) can help us achieve this, as it can observe 2400 objects simultaneously for spectral observation.

Investigating the effect of other environmental parameters such as cluster environments and clustercentric distance to AGN activity can also help shed light to the MIR AGN activity - environment relation. Therefore, studies focusing on searching cluster candidates will be beneficial to us as well. One of the pioneering studies dedicated to finding clusters \textit{AKARI} NEPW field is Huang et al. (in prep). They used the 336457 sources with local galaxy densities in this study to find clusters photometrically using a friends-of-friends (FoF) algorithm to over-densities at $z$ $\leq$ 1.1. They were able to find 88 cluster candidates with a total of 4390 cluster galaxies. Upon crossmatching with our 1210 MIR sources, only 26 of them belong to Huang et al. (in prep)'s cluster galaxy catalog. Due to the small sample size of cluster galaxies in our study, cluster-field classification was not used in this study. In addition, because these cluster candidates are found photometrically, cluster masses cannot be measured. This makes calculating clustercentric distances confusing because cluster masses are required to calculate virial radius for scaling the clustercentric distances. Failing to do so would obfuscate the physical meaning of these distances due to redshift effect \citep{Raichoor2012}. Identification of these cluster candidates is advised to confirm their nature. %Some of the instruments that can achieve this goal are the X-ray detector \textit{eROSITA} \citep{Merloni2012} and the millimeter-wave camera TolTEC (Bryan et al. 2018).}

Lastly, it is important to look at how galaxy environment affects other related galaxy properties such as stellar mass and star formation rate. Previous works have investigated the connection between galaxy environment, star formation activity, and AGN activity \citep[e.g.,][]{Sabater2013, vonderLinden2010}. In addition, many studies \citep[e.g.,][]{Lonsdale2006} have suggested that ULIRGs are advanced merger systems that fuel both star formation and AGN activity. Therefore, investigating the environmental effect on our sample's star formation is crucial to elucidate this dilemma. %This will be the focus of our future work (Santos et al. in prep).

\section{Conclusions}
\label{sec:conclusion}

We investigate how galaxy environments affect dust-obscured AGN activity with the \textit{AKARI} Infrared Telescope. We made use of \textit{AKARI}'s continuous 9-band filter coverage to provide us with a more detailed analysis of the emitted energy in mid-IR range and therefore it allows us to observe obscured AGN activities and produce numerous SEDs of galaxies which previous infrared spectrographs (e.g., \textit{Spitzer} IRS) cannot achieve. We used 1120 sources within 0 $<$ $z$ $\leq$ 1.2 from \cite{Kim2021}'s \textit{AKARI} NEPW field catalogue. This is one of the largest sample of galaxies with MIR SEDs which other spectrographs and MIR telescopes cannot do. We estimated the galaxy properties using CIGALE following \cite{Wang2020}'s parameter and module settings and \cite{Ho2021}'s photo-$z$ for sources without spec-$z$. We were able to achieve a larger amount of sources (almost one order of magnitude larger) compared to what other spectrographs (e.g., \textit{Spitzer}/IRS \citep{Wang2011}) can achieve. Our work involves disentangling the effects of redshift, environment, and total IR luminosity to shed light to the true role of galaxy environment in MIR AGN activity. Normalised local galaxy density values were calculated and an edge correction method was applied to sources near the survey edge (see Appendix \ref{sec:appendixB} for more information). We quantified the relations based on AGN contribution fraction (${\rm frac}_{\rm AGN}$) and AGN number fraction (${\rm N}_{\rm AGN}/{\rm N}_{\rm total}$). %We also made use of four different environmental parameters: normalized local density, $\Sigma_{10}^*$, cluster-field galaxy classification, cluster richness, $N_{rich}$, and clustercentric distance, ${\rm d}_{\rm cc}$. 

Our main results are as follows:

\noindent 1. Overall, AGN activity of our MIR sample do not depend on local density. However, we observed that for ULIRGs, at our highest redshift bin (0.70 < $z$ $\leq$ 1.10 and 0.80 < $z$ $\leq$ 1.20), AGN activity (i.e., ${\rm frac}_{\rm AGN}$ and ${\rm N}_{\rm AGN}/{\rm N}_{\rm tot}$) increases with density, but for our intermediate redshift bin (0.35 < $z$ $\leq$ 0.70 and 0.45 $<$ $z$ $\leq$ 0.80), AGN activity decreases with density. But these trends are not statistically significant (p $\geq$ 0.060 at the intermediate redshift bin, and p $\geq$ 0.139 at the highest redshift bin).

\noindent 2. The reversal of AGN activity-environment relationship in ULIRGs may occur at $z$ $\approx$ 0.8, although we are limited to the large uncertainties of our results.

\noindent 3. Because of the hinted reversal in the AGN activity-environment relations for ULIRGs and the lack of evidence for less luminous sources (IRGs and LIRGs), we suggest that host galaxy properties play a role in the environmental effects of MIR AGN activity, wherein the most luminous/most massive IR galaxies are the ones most affected by galaxy environment \citep[e.g.,][]{Lopes2017, Pimbblet2013, Miraghaei2020}. ULIRG populations may exhibit different properties in different epochs, which may cause the observed relation reversal at different redshifts \citep{Lonsdale2006}. Unveiling the underlying mechanisms that may explain the behavior of ULIRGs in their AGN activity-environment relation is an important direction to investigate further with upcoming space missions (e.g., \textit{JWST}, \textit{Euclid}, \textit{SPHEREx}) that will target the NEPW field.

Confirming the true nature of over-densities and possible cluster/group candidates (Huang et al. in press) are of paramount importance to further study the environments of galaxies detected in the MIR-FIR wavelengths. In addition, understanding how other galaxy properties (e.g., star formation activities) are affected by galaxy environment and its connection to our results will also be beneficial in understanding gas physics in galaxies. %will be the main focus of our next paper (Santos et al. in prep).

\section*{Acknowledgements}
We thank the anonymous referee for many insightful comments, which improved the paper.
This research is based on observations with \textit{AKARI}, a JAXA project with the participation of ESA. TG acknowledges the support by the Ministry of Science and Technology of Taiwan through grant 108-2628-M-007-004-MY3. AYLO and TH were supported by the Centre for Informatics and Computation in Astronomy (CICA) at National Tsing Hua University (NTHU) through a grant from the Ministry of Education of the Republic of China (Taiwan). AYLO’s visit to NTHU was also supported by the Ministry of Science and Technology of the ROC (Taiwan) grant 105-2119-M-007-028-MY3, kindly hosted by Prof. Albert Kong. This work used high-performance computing facilities operated by the CICA at National Tsing Hua University. This equipment was funded by the Ministry of Education of Taiwan, the Ministry of Science and Technology of Taiwan, and National Tsing Hua University. TM is supported by UNAM-DGAPA (PASPA/PAPIIT IN111319) and CONACyT Grant 252531. HSH was supported by the New Faculty Startup Fund from Seoul National University. A. Pollo and A. Poliszczuk were supported by the Polish National Science Centre grant UMO-2018/30/M/ST9/00757 and by Polish Ministry of Science and Higher Education grant DIR/WK/2018/12.
%%%%%%%%%%%%%%%%%%%%%%%%%%%%%%%%%%%%%%%%%%%%%%%%%%

\section*{Data availability}

The data used in this article are available upon request to the corresponding author. The rest of the figures underlying this article are available in its online supplementary material.



%%%%%%%%%%%%%%%%%%%%%%%%%%%%%%%%%%%%%%%%%%%%%%%%%%

%%%%%%%%%%%%%%%%% APPENDICES %%%%%%%%%%%%%%%%%%%%%

\appendix

\section{Mock Analysis in CIGALE}
\label{sec:appendixA}

To assess the reliability of our parameter settings in \texttt{CIGALE}, we made use of \texttt{CIGALE}'s mock analysis capability to produce a mock/artificial catalogue based on our photo-$z$ and spec-$z$ sources. First, \texttt{CIGALE} considers the best fit for each object in the photo-$z$/spec-$z$ source catalogue. These best fits then become the basis for the mock catalogue. Each quantity to be constrained were modified by adding a value taken from a Gaussian distribution whose standard deviation is similar to the quantity's uncertainty. The resulting artificial catalogue is then analysed in the same way as the original catalogue. Finally, the known physical properties of the mock catalogue (henceforth called the \textit{exact value}) are compared with the estimated properties from the likelihood distribution (henceforth called the \textit{estimated value}). This way, we can investigate if our constraints on the galaxy properties are reliable or not. We focus on the mock analysis of AGN contribution fraction (${\rm frac}_{\rm AGN}$) and star formation rate (SFR).

\begin{figure}
	\includegraphics[width=\columnwidth]{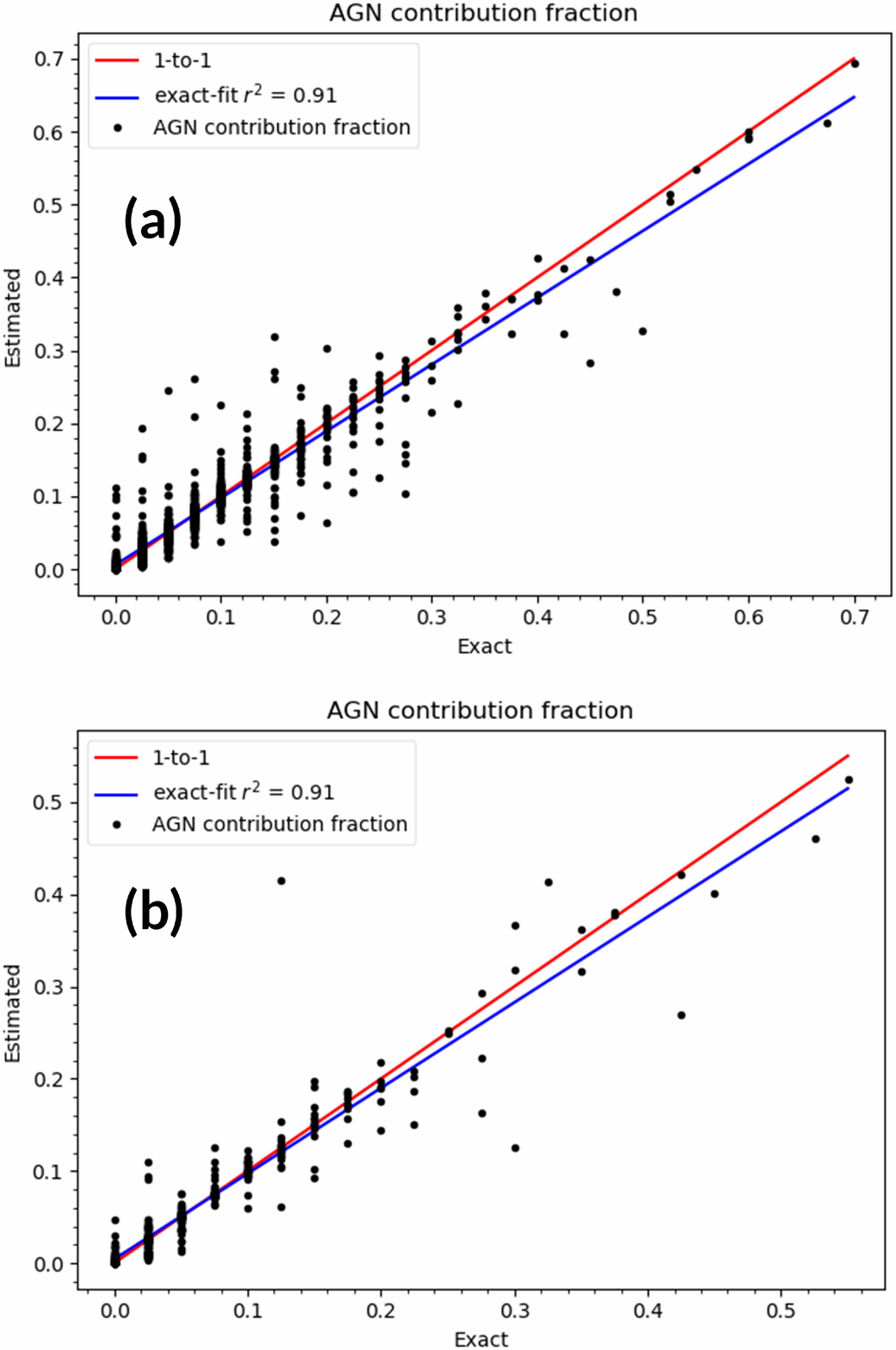}  
    \caption{Estimated AGN contribution fraction vs. exact AGN contribution fraction produced by mock analysis of (a) the 1385 photo-$z$ sources, and (b) 409 spec-$z$ sources. The red and black lines refer to the best fit line via linear regression and the 1:1 line, respectively. The goodness-of-fit coefficient of the best fit line is $r^2$ =  0.91.}
    \label{fig:mock_agn}
\end{figure}

\begin{figure}
	\includegraphics[width=\columnwidth]{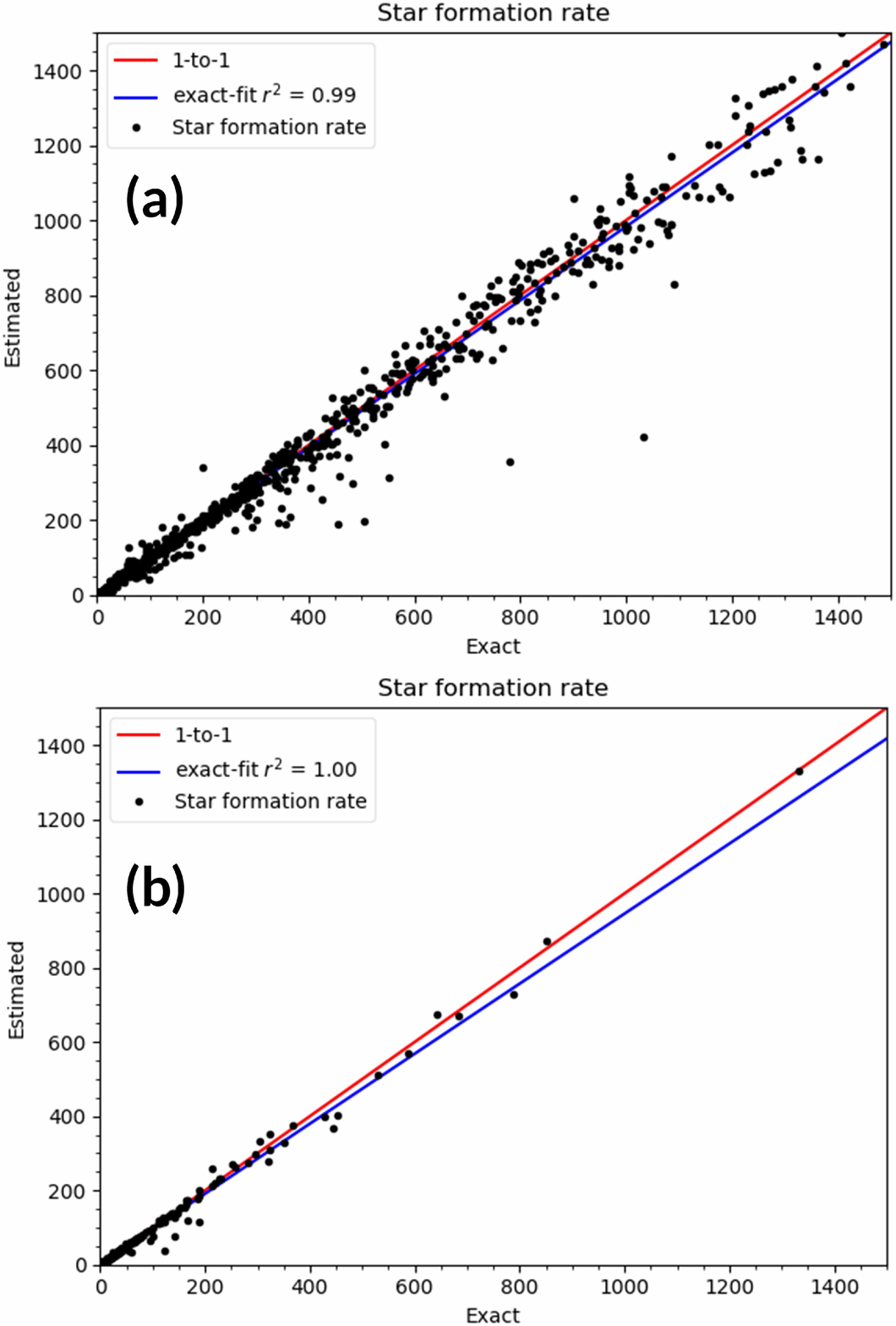}  
    \caption{Same as Fig.~\ref{fig:mock_agn} but for star formation rate. The goodness-of-fit coefficient of the best fit line is $r^2$ = 1.00}
    \label{fig:mock_sfr}
\end{figure}

Figures~\ref{fig:mock_agn} and~\ref{fig:mock_sfr} show the results of our mock analysis for ${\rm frac}_{\rm AGN}$ and SFR, respectively. We also show the mock analysis results using only the 1385 photo-$z$ sources (a), and 409 spec-$z$ sources (b). It is clear from our mock analysis results that we were able to reliably constrain these properties, as their goodness-of-fit coefficient (or coefficient of determination) are 0.91 for ${\rm frac}_{\rm AGN}$, and about 1.00 for SFR.

\section{Robustness of Edge Correction for Density Calculation}
\label{sec:appendixB}

The robustness of this edge correction method was tested on a subset consisting of HSC galaxies in the \textit{AKARI} NEPW field contained in a circular area centred at the centre of the survey with a radius of 30$^\prime$. The galaxies included in this area are very far from the survey edge (at least $\sim$ 120$^\prime$ away from the survey edge), thus their actual density values are very likely to be correct. Similar selection criteria with the initial density calculation was also applied to the subset. 57834/366791 ($\sim$ 15.8\%) galaxies were selected as the subset galaxies. We used n = 2,3,5,10,15, and 20 in our comparison to check the effectiveness of our edge correction method for varying values of n. For each value of n, the new edge distances and densities of each source were calculated again. Edge subset galaxies were selected using the same criteria as before, and then we correct their density values. 
In this part, the recalculated density values of the sources are called "uncorrected density" values, while their edge-corrected density values (for edge galaxies) are called "corrected density" values. These values are compared with their "true density" values, which are their original density values when we use all the 366791 sources for density calculation.

%\begin{table}
%\caption{\label{tab:edgegalaxies}Number and percentage of edge galaxies for varying values of n using our subset sample.}
%\begin{center}
%\begin{ruledtabular}
%\begin{tabular}{ccc}
%n & \makecell{Number of \\ edge galaxies} & \makecell{Percentage of \\ edge galaxies} \\ \hline
%2 & 1418 & 2.45\% \\
%3 & 1771 & 3.06\% \\
%5 & 2359 & 4.08\% \\
%10 & 3369 & 5.83\% \\
%15 & 4162 & 7.20\% \\
%20 & 4805 & 8.31\% 
%\end{tabular}
%\end{ruledtabular}
%\end{center}
%\end{table}

\begin{figure}
	\includegraphics[width=0.8\columnwidth]{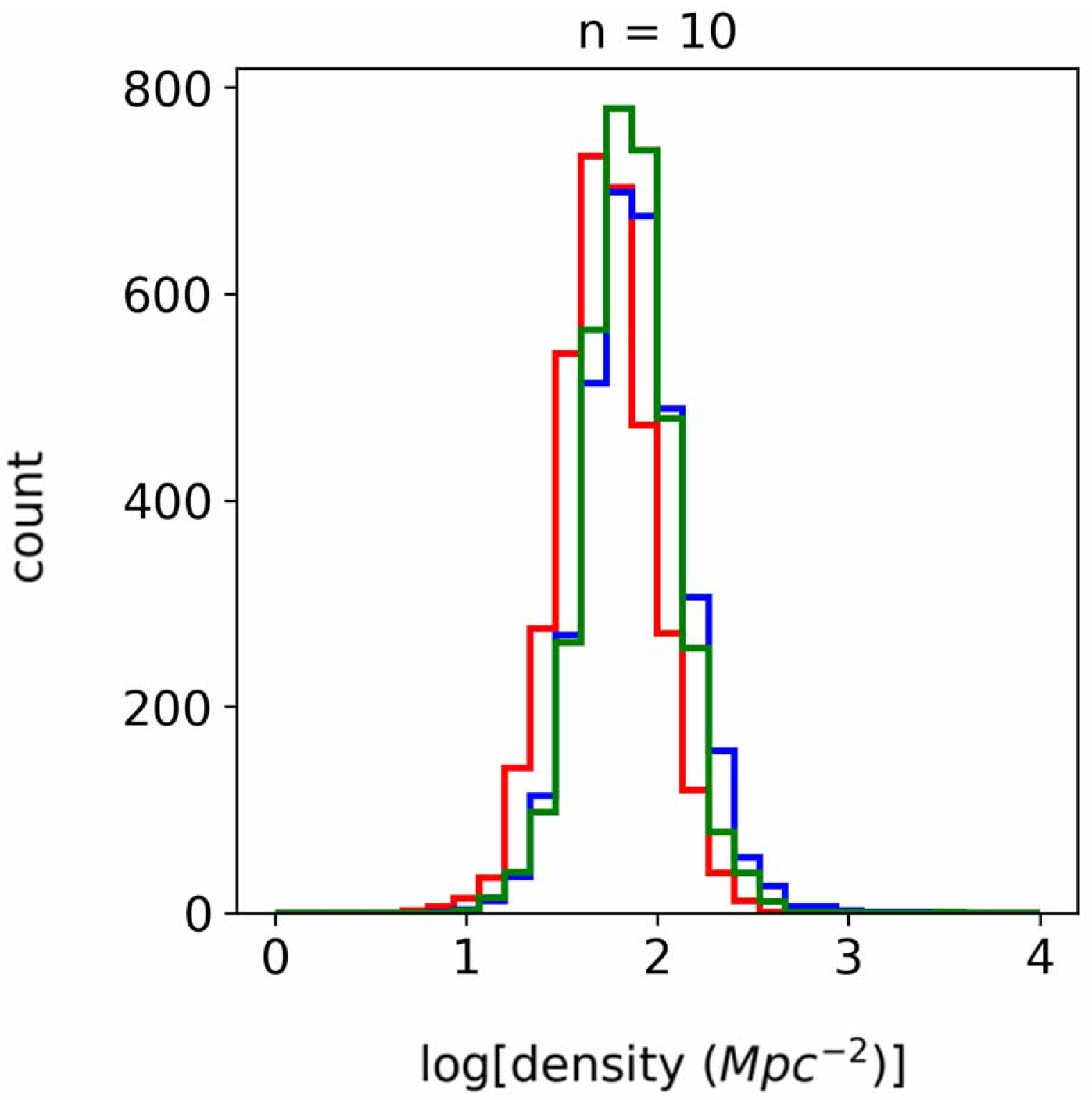}  
    \caption{Histogram of uncorrected (red histogram), corrected (blue histogram), and true (green histogram) densities in ${\rm Mpc}^{-2}$ of edge subset galaxies for n=10. For n=10, there are 3369/57834 (5.83\%) subset galaxies that are edge galaxies.}
    \label{fig:edgecorrection_varying_n}
\end{figure}

Fig.~\ref{fig:edgecorrection_varying_n} shows the histogram of uncorrected (red histogram), corrected (blue histogram), and true (green histogram) densities of edge galaxies for n=10 for our subset sample. Note that the number of edge galaxies changes for each value of n (see Table S1 in the Supplementary Material).

At very low values of n (i.e., n = 2 and 3), the blue (corrected densities) and green (true densities) histograms do not overlap with each other, indicating that our edge correction does not work well with lower values of n. Thus, we cannot use small values of n in our research. At n = 5, 10, 15, and 20, the histograms of corrected and true densities overlap well, signifying that our edge correction method works well in these values of n. However, larger n values (i.e., n = 15, 20) might cause the calculated densities to be insensitive to overdensities (that is, if we use larger values of n, we may miss possible overdense regions with galaxy members less than n). Therefore, we chose n = 10 in this work which is a compromise between the effectiveness of edge correction and sensitivity of galaxy density in finding overdensities. The histogram of uncorrected, corrected, and true densities for other values of n are shown in the Supplementary Material.

\begin{figure*}
	\includegraphics[width=1.6\columnwidth]{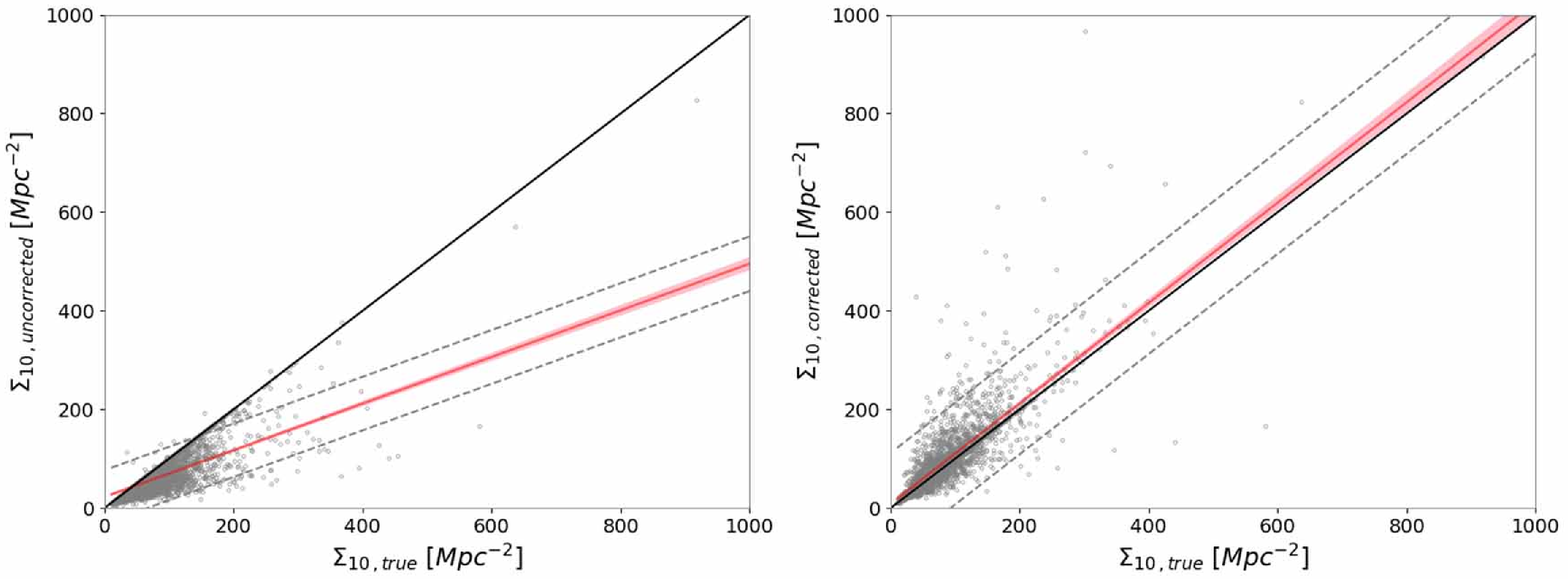}  
    \caption{Comparison between the uncorrected density, $\Sigma_{\rm n, uncorrected}$, vs.  corrected density, $\Sigma_{\rm n, corrected}$ (left column) and uncorrected density, $\Sigma_{\rm n,  uncorrected}$, vs. true density, $\Sigma_{\rm n, true}$ (right column) for n=10. The red, black, and dashed lines refer to the best fit line of the sources (gray points), 1:1 line, and 95\% prediction interval, respectively. The red shaded region is the 95\% confidence limits of the best fit line.}
    \label{fig:bestfit}
\end{figure*}

Figure~\ref{fig:bestfit} shows the comparison between the uncorrected density, $\Sigma_{\rm uncorrected}$ and corrected density, $\Sigma_{\rm corrected}$ (left column) versus true density, $\Sigma_{\rm true}$ (right column) for n=10. Our edge correction method works best at n=10; using a lower value of n gives an overestimated density after edge correction, while a higher value of n makes cluster finding less sensitive to overdensities, not to mention edge correction does not also work well with them. With n=10, the edge corrected density values are approximately close to the actual density values of edge galaxies. The rest of the plots for other values of n are shown in the Supplementary Material.

%\begin{figure*}
%	\includegraphics[width=1.8\columnwidth]{uncorrected_vs_corrected_plots2.eps}  
%    \caption{Continuation of Fig.~\ref{fig:bestfit}}
%    \label{fig:bestfit2}
%\end{figure*}

%%%%%%%%%%%%%%%%%%%%%%%%%%%%%%%%%%%%%%%%%%%%%%%%%%

% Don't change these lines
\bsp	% typesetting comment
\label{lastpage}
\end{document}